\documentclass[twocolumn,usenatbib]{aastex62}

\usepackage{amsfonts,amssymb,amsbsy,amsmath}
\usepackage{mathtools}
\usepackage{upgreek}

\usepackage{natbib}
\usepackage{graphicx}
\usepackage{xifthen}
\usepackage{bigints}
\usepackage{booktabs}
\usepackage{nicefrac}
\usepackage{color}
\usepackage{url}
\usepackage{hyperref} 
\usepackage[theorems]{tcolorbox}

\catcode`_=\active
\newcommand_[1]{\ensuremath{\sb{\mathrm{#1}}}}
\newcommand*{\logten}{\mathop{\log_{10}}}   

\newcommand{\cequiv}[1]{\mathrel{\stackrel{\makebox[0pt]{\mbox{\normalfont\tiny #1}}}{\equiv}}}
\newcommand{\cpropto}[1]{\mathrel{\stackrel{\makebox[0pt]{\mbox{\normalfont\tiny #1}}}{\propto}}}
\newcommand{\newpar}{{}}

\newcommand{\pdfbig}[1]{{\pi\big(#1\big)}}

\newcommand{\up}{\operatorname}
\newcommand{\diff}{{\up{d}}}

\newcommand{\bs}{\boldsymbol}

\newcommand{\Mean}{{\bs{\mathrm{\mu}}}}

\newcommand{\CovMat}{{\bs{\mathrm{\Sigma}}}}

\newcommand{\numChar}{{n}} 

\newcommand{\paraChar}{{p}}

\newcommand{\episChar}{{n}}

\newcommand{\npe}[1][]{{ \numChar_{ \ifthenelse{\isempty{#1}}{\paraChar\episChar}{{\paraChar\episChar,#1}} } }} 

\newcommand{\epis}{{nois}}

\def\gtrsim{\mathrel{\hbox{\rlap{\hbox{\lower4pt\hbox{$\sim$}}}\hbox{$>$}}}}
\def\lessim{\mathrel{\hbox{\rlap{\hbox{\lower4pt\hbox{$\sim$}}}\hbox{$<$}}}}

\newcommand{\rmz}{{\rm z}}

\newcommand{\emodel}[1][]{{ M_{ \ifthenelse{\isempty{#1}}{\epis}{{\epis,#1}} } }}
\newcommand{\emodelz}[1][]{{ M_{ \rmz\ifthenelse{\isempty{#1}}{\epis}{{\epis,#1}} } }}
\newcommand{\emodellgrb}[1][]{{ M^\lgrb_{ \ifthenelse{\isempty{#1}}{\epis}{{\epis,#1}} } }}

\newcommand{\observed}{{obs}}
\newcommand{\intrinsic}{{int}}

\newcommand{\censored}{{cen}}
\newcommand{\cosmic}{{tru}}

\newcommand{\eff}{{eff}}
\newcommand{\obsi}{{obs,i}}
\newcommand{\inti}{{int,i}}
\newcommand{\lgrb}{{\rm g}}
\newcommand{\model}{{\mathcal{R}}}
\newcommand{\mint}{{\model_\cosmic}} 
\newcommand{\mobs}{{\model_\censored}} 
\newcommand{\mintlgrb}{{\model_\cosmic^\lgrb}}
\newcommand{\meff}{{\eta_\eff}}

\newcommand{\param}{{\bs{\theta}}}
\newcommand{\pobs}{{\param_\censored}} 
\newcommand{\pint}{{\param_\cosmic}} 
\newcommand{\peff}{{\param_{eff}}}
\newcommand{\pintlgrb}{{\param_\cosmic^\lgrb}}
\newcommand{\pz}{{\param_z}}

\newcommand{\eparam}[1][]{{ \param_{ \ifthenelse{\isempty{#1}}{\epis}{{\epis,#1}} } }}
\newcommand{\eparamz}[1][]{{ \bs\param^\rmz_{ \ifthenelse{\isempty{#1}}{\epis^\rmz}{{\epis,#1}} } }}
\newcommand{\eparamlgrb}[1][]{{ \bs\param^\lgrb_{ \ifthenelse{\isempty{#1}}{\epis^\lgrb}{{\epis,#1}} } }}

\newcommand{\domain}{{ \up{\Omega} }}
\newcommand{\domaindint}{{ \domain(\dint) }}

\newcommand{\domaindsetintlgrb}{{ \domain(\dsetintlgrb) }}
\newcommand{\domainpobs}{{ \domain(\pobs) }}

\newcommand{\data}{{\bs{D}}}

\newcommand{\dint}{{\data_\intrinsic}} 
\newcommand{\dintp}{{\data^{\possible}_\intrinsic}} 
\newcommand{\dinti}{{\data_\inti}} 

\newcommand{\dobsilgrb}{{\data^\lgrb_\obsi}}
\newcommand{\dobsilgrbmode}{{{\widehat\data}^\lgrb_\obsi}}

\newcommand{\dintlgrb}{{\data^\lgrb_\intrinsic}}
\newcommand{\dintilgrb}{{\data^\lgrb_\inti}}

\newcommand{\truth}{{\bs{R}}}
\newcommand{\possible}{{*}}
\newcommand{\truthset}{{\mathcal{R}}}

\newcommand{\truthsubset}[1][]{{ \truthset_{ \ifthenelse{\isempty{#1}}{\truth}{{\truth_{#1}}} } }}
\newcommand{\ptruthsubset}[1][]{{ \truthset_{ \ifthenelse{\isempty{#1}}{\truth}{{\truth_{#1}}} }^\possible }}

\newcommand{\dset}{{\mathcal{D}}}
\newcommand{\dsetintlgrb}{{\dset^\lgrb_\intrinsic}}
\newcommand{\dsetobslgrb}{{\dset^\lgrb_\observed}}
\newcommand{\zset}{{\mathcal{Z}}}
\newcommand{\nint}{{N_{int}}}
\newcommand{\nobs}{{N_{obs}}}

\renewcommand*{\thefootnote}{\fnsymbol{footnote}}

\renewcommand*{\thefootnote}{\fnsymbol{footnote}}

\newcommand{\xx}[1][]{{ \ifthenelse{\isempty{#1}}{\textcolor{red}{XXX}}{\textcolor{red}{~(XXX {#1} XXX)~}} }}

\newcommand{\liso}{{L_{iso}}}
\newcommand{\eiso}{{E_{iso}}}
\newcommand{\epkz}{{E_{pz}}}
\newcommand{\durz}{{T_{90z}}}
\newcommand{\pbol}{{P_{bol}}}
\newcommand{\sbol}{{S_{bol}}}
\newcommand{\epk}{{E_{p}}}
\newcommand{\dur}{{T_{90}}}
\newcommand{\pph}{{P_{ph}}}
\newcommand{\zi}{{z_{i}}}
\newcommand{\lisoi}{{L_{iso,i}}}
\newcommand{\eisoi}{{E_{iso,i}}}
\newcommand{\epkzi}{{E_{pz,i}}}
\newcommand{\durzi}{{T_{90z,i}}}

\newcommand{\pbolimode}{{\widehat{P}_{bol,i}}}
\newcommand{\sbolimode}{{\widehat{S}_{bol,i}}}
\newcommand{\epkimode}{{\widehat{E}_{p,i}}}
\newcommand{\durimode}{{\widehat{T}_{90,i}}}

\newcommand{\ldis}{{d_L}}
\newcommand{\thresh}{{th}}
\newcommand{\threshM}{{\mu_\thresh}}
\newcommand{\threshS}{{\sigma_\thresh}}
\newcommand{\mz}{{\dot\zeta}}

\defcitealias{hopkins2006normalization}{H06}
\defcitealias{li2008star}{L08}
\defcitealias{butler2010cosmic}{B10}
\defcitealias{madau2014cosmic}{M14}
\defcitealias{madau2017radiation}{M17}
\defcitealias{fermi2018gamma}{F18}

\graphicspath{{./}{figures/}}

\shorttitle{BATSE LGRBs redshift catalog}
\shortauthors{Osborne, Shahmoradi, \& Nemiroff}

\begin{document}

\title{A Multilevel Empirical Bayesian Approach to Estimating the Unknown Redshifts of 1366 BATSE Catalog Long-Duration Gamma-Ray Bursts}


\email{
joshua.osborne@mavs.uta.edu (JAO) \\
shahmoradi@utexas.edu (AS) \\
nemiroff@mtu.edu (RJN) \\
}

\author{Joshua A. Osborne}
\affiliation{
Department of Physics \\
The University of Texas \\
Arlington, TX 76010, USA
}

\author{Amir Shahmoradi}
\affiliation{
Department of Physics \\
The University of Texas \\
Arlington, TX 76010, USA
}
\affiliation{
Data Science Program, College of Science \\
The University of Texas \\
Arlington, TX 76010, USA
}

\author{Robert J. Nemiroff}
\affiliation{
Department of Physics \\
Michigan Technological University \\
Houghton, MI 49931, USA
}

\begin{abstract}
    We present a catalog of the probabilistic redshift estimates and for 1366 individual Long-duration Gamma-Ray Bursts (LGRBs) detected by the Burst And Transient Source Experiment (BATSE). This result is based on a careful selection and modeling of the population distribution of 1366 BATSE LGRBs in the 5-dimensional space of redshift and the four intrinsic prompt gamma-ray emission properties: the isotropic 1024ms peak luminosity (\liso), the total isotropic emission (\eiso), the spectral peak energy (\epkz), as well as the intrinsic duration (\durz), while carefully taking into account the effects of sample incompleteness and the LGRB-detection mechanism of BATSE. Two fundamental plausible assumptions underlie our purely-probabilistic approach: 1. LGRBs trace, either exactly or closely, the Cosmic Star Formation Rate and 2. the joint 4-dimensional distribution of the aforementioned prompt gamma-ray emission properties is well-described by a multivariate log-normal distribution.
    Our modeling approach enables us to constrain the redshifts of individual BATSE LGRBs to within $0.36$ and $0.96$ average uncertainty ranges at $50\%$ and $90\%$ confidence levels, respectively. Our redshift predictions are completely at odds with the previous redshift estimates of BATSE LGRBs that were computed via the proposed phenomenological high-energy relations, specifically, the apparently-strong correlation of LGRBs' peak luminosity with the spectral peak energy, lightcurve variability, and the spectral lag. The observed discrepancies between our predictions and the previous works can be explained by the strong influence of detector threshold and sample-incompleteness in shaping these phenomenologically-proposed high-energy correlations in the literature.
\end{abstract}


\section{Introduction}
\label{sec:introduction}

    Throughout almost a decade of continuous operation, the Burst And Transient Source Experiment (BATSE) onboard the now-defunct Compton Gamma-Ray Observatory \citep{meegan1992spatial} detected more than $2700$ Gamma-Ray Bursts (GRBs). The BATSE catalog of GRBs provided the first solid evidence for the existence of at least two classes of GRBs: the short-hard (SGRBs) and the long-soft (LGRBs) \citep[e.g.,][]{kouveliotou1993identification}.\newpar

    Traditionally, new GRB events have been classified into one of the two classes based on a sharp cutoff on the bimodal distribution of the observed duration ($\dur$) of the prompt gamma-ray emission, generally set to $\dur\sim2-3[s]$. However, the dependence of the observed duration of GRBs on the gamma-ray energy and the detector's specifications \citep[e.g.,][]{fenimore1995gamma, nemiroff2000pulse, qin2012comprehensive} has prompted many studies in search of less-biased alternative methods of GRB classification, typically based on a combination of the prompt gamma-ray and afterglow emissions as well as the host galaxy's properties \citep[e.g.,][]{gehrels2009gamma, zhang2009discerning, shahmoradi2010hardness, goldstein2011new, shahmoradi2011possible, zhang2012revisiting, shahmoradi2013multivariate, shahmoradi2013gamma, shahmoradi2014classification, shahmoradi2015short, lu2014amplitude} or based on the prompt-emission spectral correlations in conjunction with the traditional method of classification \citep[e.g.,][]{qin2013statistical}.\newpar


    Ideally, the classification of GRBs should be independent of their cosmological distances from the earth and free from potential sample biases due to detector specifications, selection effects, sample incompleteness, and should solely rely on their intrinsic properties. Such classification methods are still missing in the GRB literature and hard to devise, mainly due to the lack of a homogenously-detected, sufficiently-large catalog of GRBs with measured redshifts.\newpar

    Several studies have already attempted to estimate the unknown redshifts of GRBs based on the apparently-strong phenomenological correlations observed between some of the spectral and temporal prompt gamma-ray emission properties of GRBs. The most prominent class of such relations are the apparently-strong correlations of the intrinsic brightness measures of the prompt gamma-ray emission (e.g., the total isotropic emission, $\eiso$, and the peak $1024ms$ luminosity, $\liso$) with other spectral or temporal properties of GRBs, such as {\it hardness} as measured by the intrinsic spectral peak energy $\epkz$ \citep[e.g.,][]{yonetoku2004gamma, yonetoku2014short}, light-curve variability \citep[e.g.,][]{fenimore2000redshifts, reichart2001possible}, the spectral lag \citep[e.g.,][]{schaefer2001redshifts}, or based on a combination of such relationships \citep[e.g.,][]{xiao2009estimating,dainotti2019gamma}.\newpar

    These methods, however, can lead to incorrect or highly biased estimates of the unknown redshifts of GRBs if the observed high-energy correlations are constructed from a small sample of GRBs (typically the brightest events) with measured redshifts. Such small samples are often collected from multiple heterogeneous surveys and may neither represent the entire population of observed GRBs (with or without measured redshift) nor represent the unobserved cosmic population. More importantly, the potential effects of detector threshold and sample-incompleteness on them are poorly understood. Such biases manifest themselves in redshift estimates that are inconsistent with estimates from other methods, examples of which have been already reported by several authors \citep[e.g.,][]{guidorzi2005testing, ashcraft2007there, rizzuto2007testing, bernardini2014comparing}.\newpar

    The selection effects in the detection, analysis, and redshift measurements of GRBs and their potential effects on the observed phenomenological high-energy correlations has been already extensively studied individually, in isolation from other correlations, \citep[e.g.,][]{petrosian1996fluence, lloyd1999distribution, petrosian1999cosmological, lloyd2000cosmological, hakkila2003sample, band2005testing, nakar2004outliers, butler2007complete, ghirlanda2008peak, nava2008peak, shahmoradi2009real, butler2009generalized, butler2010cosmic, shahmoradi2011cosmological, shahmoradi2011possible, shahmoradi2013gamma, dainotti2015selection, petrosian2015cosmological}. However, an ultimate resolution to the problem of estimating the unknown redshifts of GRBs in catalogs requires simultaneous multidimensional modeling of the intrinsic population distribution of GRB attributes, subject to the effects of detector threshold and sample incompleteness on their joint observed distribution \citep[e.g.,][]{butler2010cosmic, shahmoradi2013multivariate, shahmoradi2014classification, shahmoradi2015short}.\newpar

    Building upon our previous studies in \citet{shahmoradi2013gamma,shahmoradi2013multivariate,shahmoradi2015short}, and motivated by the existing gap in the knowledge of the redshifts of LGRBs in BATSE catalog \citep[e.g.,][]{paciesas1999fourth, goldstein2013batse}, which as of 2020, constitutes the largest catalog of homogenously-detected GRBs, here we present a methodology and modeling approach to constraining the redshifts of 1366 BATSE LGRBs. Despite lacking complete knowledge of the true cosmic rate and redshift distribution of LGRBs, we show that it is possible to constrain the redshifts of individual BATSE LGRBs to within uncertainty ranges of width $0.7$ and $1.7$, on average, at $50\%$ and $90\%$ confidence levels, respectively. Our methodology relies on two plausible assumptions which are strongly supported by the currently existing evidence: 1. LGRBs trace the cosmic Start Formation Rate (SFR) or a metallicity-corrected SFR \citep[e.g.,][]{butler2010cosmic,pontzen2010nature} and, 2. the joint distribution of the four main prompt gamma-ray-emission properties of LGRBs is well described by a multivariate log-normal distribution \citep[e.g.,][]{shahmoradi2013multivariate, shahmoradi2015short}. The presented work also paves the way towards a detector-independent minimally-biased phenomenological classification method for GRBs solely based on the intrinsic prompt gamma-ray data of individual events.\newpar

    In the following sections, we present an attempt to further uncover some of the tremendous amounts of useful, yet unexplored information that is still buried in this seemingly archaic catalog of GRBs. Towards this, we devote section \S\ref{sec:methods} of this manuscript to the development of redshift inference methodology, which includes a discussion of the data collection procedure in \S\ref{sec:methods:data}, a generic description of our probabilistic modeling approach via a toy problem in \S\ref{sec:methods:bayesianToyProblem}, followed by detailed descriptions of the proposed methodology for estimating redshifts in \S\ref{sec:methods:redshiftEstimation}, the cosmic SFR assumptions underlying our model in \S\ref{sec:methods:lgrbRateDensity}, the construction of an LGRB world model in \S\ref{sec:methods:lgrbWorldModel}, and a review of the BATSE LGRB detection algorithm and our approach to modeling the BATSE LGRBs sample incompleteness in \S\ref{sec:methods:modelingSampleIncompleteness}. The predictions of the model are presented in \S\ref{sec:results}, followed by a discussion of the implications of the results, comparison with previous independent redshift estimates, and possible reasons for the observed discrepancies between the results of this study and the previous studies in \S\ref{sec:discussion}.

\section{Methods}
\label{sec:methods}

\subsection{BATSE LGRB Data}
\label{sec:methods:data}

    Fundamental to our inference problem is the issue of obtaining a dataset of BATSE LGRBs that is minimally-biased and representative of the population distribution of LGRBs detectable by the BATSE Large Area Detectors (LADs). The traditional method of GRB classification based on a sharp cutoff line on the observed duration variable $\dur$ set at $2-3 [s]$ \citep[][]{kouveliotou1993identification} has been shown insufficient for an unbiased classification since the duration distributions of LGRBs and SGRBs have significant overlap \citep[e.g.,][]{butler2010cosmic, shahmoradi2013multivariate, shahmoradi2015short}. Instead, we follow the multivariate fuzzy classification approach of \citet{shahmoradi2013multivariate, shahmoradi2015short} to segregate the two populations of BATSE LGRBs and SGRBs based on their estimated observed spectral peak energies ($\epk$) from \citet{shahmoradi2010hardness} and $\dur$ from the current BATSE catalog \citep[][]{goldstein2013batse}. This leads us to a sample of 1366 LGRBs versus 565 SGRBs in the current BATSE catalog. We refer the interested reader to \citet{shahmoradi2013multivariate, shahmoradi2015short} for extensive details of the classification procedure.\newpar

    For our analysis, we also compute, as detailed in \cite{shahmoradi2013multivariate, shahmoradi2015short}, the 1024 [ms] bolometric peak flux ($\pbol$) and the bolometric fluence (\sbol) of these events from the current BATSE catalog for inclusion in the analysis, in addition to $\epk$ and $\dur$. Taking into account the measurement uncertainties associated with each BATSE event, we can therefore, represent the $i$th event, $\dobsilgrb$, in the BATSE catalog by an a-priori `known' measurement uncertainty model, $\emodellgrb[i]$, that together with its `known' parameters, $\eparamlgrb[i]$, determine the joint 4-dimensional probability density function of the observed attributes of the event,
    \begin{eqnarray}
        \label{eq:dobsilgrb}
        \pi\big( \dobsilgrb &|& \emodellgrb[i] , \eparamlgrb[i] \big)  \nonumber \\
                            &\propto& \emodellgrb[i] \big( \dobsilgrb , \eparamlgrb[i] \big) ~.
    \end{eqnarray}

    The modes of these uncertainty models are essentially the values reported in the BATSE catalog and \citet{shahmoradi2010hardness},
    \begin{equation}
        \label{eq:dobsilgrbmode}
        \dobsilgrbmode = [\pbolimode,\sbolimode,\epkimode,\durimode] ~,
    \end{equation}

    The entire BATSE dataset of 1366 LGRB events attributes is then represented by the collection of pairs of such uncertainty models and their parameters,
    \begin{equation}
        \label{eq:dsetobslgrb}
        \dsetobslgrb = \big\{ \big( \emodellgrb[i] , \eparamlgrb[i] \big) : ~1\leq i\leq1366 \big\} ~,
    \end{equation}

    Note that throughout this manuscript, the appearance of `$\lgrb$' as a superscript solely indicates that the quantity relates to or depends on the four main {\it physical} LGRB properties considered in this study, {\it excluding} any redshift ($z$) information.\newpar

    The BATSE catalog observations appear to have been reported with the assumption that the measurement uncertainty models for all events are multivariate Normal distributions, with their mean vectors being the values reported in the catalog, and their covariance matrices being diagonal, with the diagonal elements representing the square of the 1-$\sigma$ errors reported in the catalog.\newpar

    \begin{figure}
        \centering
        \includegraphics[width=0.48\textwidth]{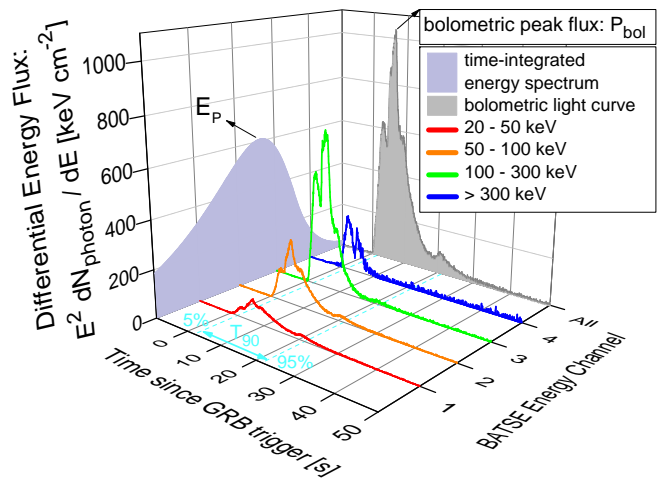}
        \caption{An example GRB lightcurve (BATSE trigger 1085) illustrating the four main LGRB properties used as input observational data in this study (Eqn. \eqref{eq:dobsilgrbmode}): the observed 1-sec bolometric peak energy flux ($\pbol$), the observed total energy fluence ($\sbol$), the observed spectral peak energy ($\epk$), and the observed duration within which the GRB event releases $90\%$ of its gamma-ray emission ($\dur$). The four red, orange, green, and blue colored lines represent the total energy flux received, as a function of time, in each of the four BATSE main energy channels.
        \label{fig:methods:data:3d}}
    \end{figure}

    In sum, our observational data is comprised of the four main prompt gamma-ray emission properties of LGRBs as described by Eqn. \eqref{eq:dobsilgrb} \& \eqref{eq:dobsilgrbmode} and illustrated by Figure \ref{fig:methods:data:3d}. Although, $\pbol$ as in \eqref{eq:dobsilgrbmode} is a conventionally-defined measure of the peak brightness and peripheral to and highly correlated with the more fundamental GRB attribute, $\sbol$, its inclusion in our GRB world model is essential as it determines, together with $\epk$, the peak {\it photon} flux, $\pph$, in $50-300$ [keV] range, based upon which BATSE LADs generally triggered on LGRBs.

\subsection{The Multilevel Empirical Bayesian Approach}
\label{sec:methods:bayesianToyProblem}

    Except a handful of GRB events, the entire BATSE catalog of GRBs lack any redshift or distance information. The knowledge of individual redshifts is absolutely necessary for accurate cosmographic studies of LGRBs. Essentially, our missing-redshift-data problem for each LGRB reduces to a set of four equations,
    \begin{align}
        \label{eq:obsIntMap}
        \liso &= 4\pi\times\ldis(z)^2\times\pbol &, \nonumber \\
        \eiso &= 4\pi\times\ldis(z)^2\times\sbol / (z+1) &, \nonumber \\
        \epkz &= \epk\times(z+1) &, \nonumber \\
        \durz &= \dur / (z+1)^\alpha &,
    \end{align}

    \noindent that exactly determine the intrinsic properties of LGRBs: the 1024 [ms] isotropic peak luminosity ($\liso$), the total isotropic emission ($\eiso$), the intrinsic spectral peak energy ($\epkz$), and the intrinsic duration ($\durz$). These four properties are collectively represented by,
    \begin{equation}
        \label{eq:dintlgrb}
        \dintilgrb = [\lisoi,\eisoi,\epkzi,\durzi] ~.
    \end{equation}

    The term $\ldis(z)$ in \eqref{eq:obsIntMap} represents the luminosity distance,
    \begin{equation}
        \label{eq:ldis}
        \ldis(z)=\frac{C}{H_0}(1+z)\int^{z}_{0}dz'\bigg[(1+z')^{3}\Omega_{M}+\Omega_{\Lambda}\bigg]^{-1/2} ~,
    \end{equation}

    \noindent and the exponent $\alpha=0.66$ in the mapping of $\dur$ to $\durz$ takes into account the cosmological time-dilation as well as an energy-band correction (i.e., K-correction) of the form $(1+z)^{-0.34}$ to the observed durations \citep[e.g.][]{gehrels2006new}. Throughout this work we assume a flat $\Lambda$CDM cosmology, with parameters set to $h=0.70$, $\Omega_M=0.27$ and $\Omega_\Lambda=0.73$ \citep{jarosik2011seven}. Also, the parameters $C$ \& $H_{0}=100h$ [km/s/MPc] stand for the speed of light and the Hubble constant respectively. The equations of \eqref{eq:obsIntMap} can be linearized by taking of the logarithms of both sides, leading to,
    \begin{align}
        \label{eq:obsIntMapLinearized}
        \log(\liso) &= \log(\pbol) + 2\log(\ldis) + \log(4\pi) , \nonumber \\
        \log(\eiso) &= \log(\sbol) + 2\log(\ldis) + \log(4\pi) - \log(z+1) , \nonumber \\
        \log(\epkz) &= \log(\epk)  + \log(z+1) , \nonumber \\
        \log(\durz) &= \log(\dur)  - \alpha\log(z+1) .
    \end{align}

    More concisely, we can write the four equations of \eqref{eq:obsIntMapLinearized} for the i$^\mathrm{th}$ GRB event as a single equation,
    \begin{equation}
        \log(\dintilgrb) = \log(\dobsilgrb) + \mathcal{F}(z_i) ~, \label{eq:obsIntMapLinearizedVectorized}
    \end{equation}

    \noindent where $\mathcal{F}(z_i)$ describes the mapping from the observer to the rest frame of the GRB given its known redshift, $z_i$. This simple equation deserves some deliberations. Despite its simplicity, this single equation is algebraically unsolvable for almost any BATSE catalog GRB as it contains two unknowns ($\dintilgrb$, $z_i$).\newpar

    Although \eqref{eq:obsIntMapLinearizedVectorized} is algebraically degenerate, the two unknowns of the equation can still be constrained probabilistically. To understand how this is possible, consider the following simple toy problem.\newpar

    \begin{figure}
        \centering
        \includegraphics[width=0.45\textwidth]{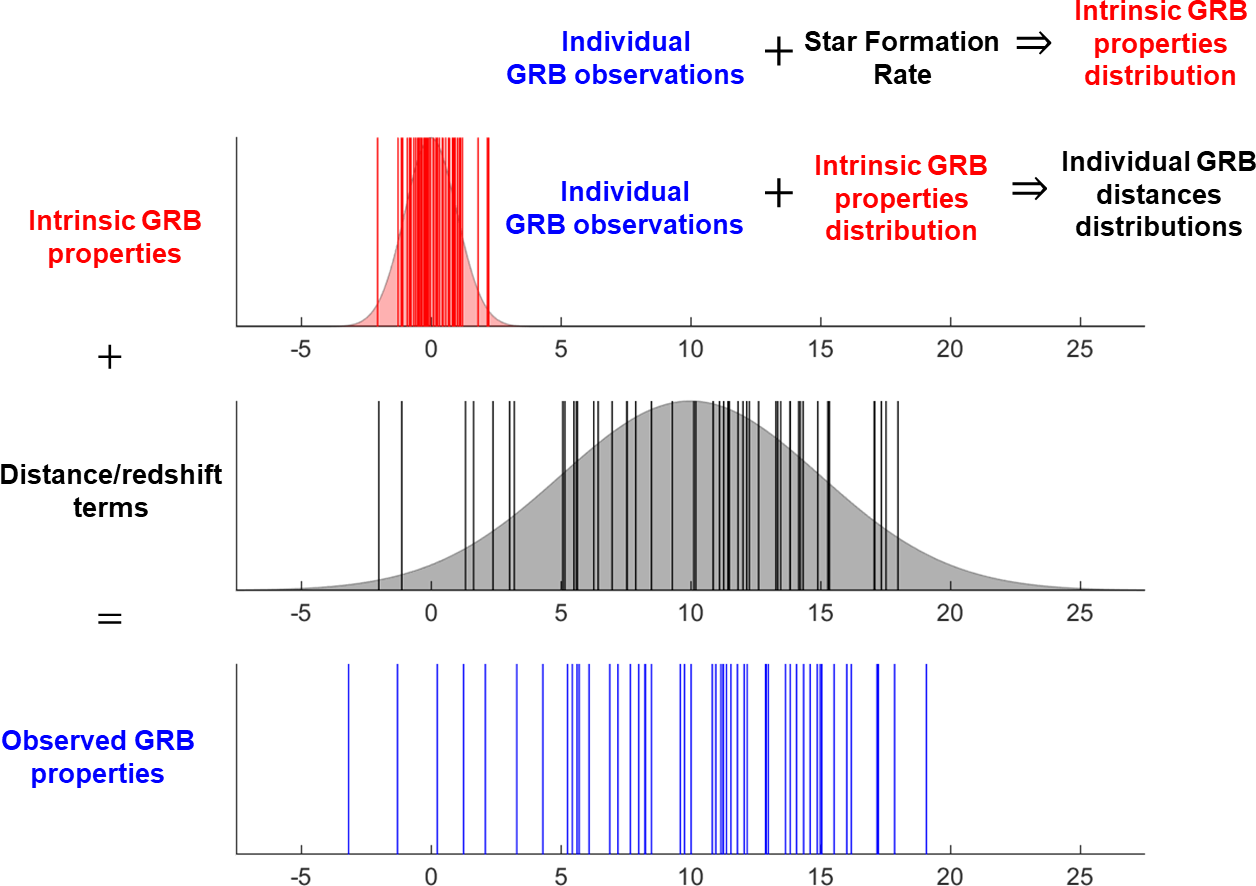}
        \caption{An schematic illustration of the multilevel empirical Bayesian approach to estimating the unknown redshifts of BATSE LGRBs (the {\color{gray}{gray lines}}). The process of inferring the redshifts involves two separate steps. At the first stage, we convolve the observed BATSE LGRB properties (the individual {\color{blue}{blue lines}}) with prior knowledge of the overall cosmic redshift distribution of LGRBs (the {\color{gray}{gray distribution}}) to infer the unknown parameters of the joint population distribution of the intrinsic properties of LGRBs (the {\color{red}{red distribution}}). Then, we combine the inferred best-fit model for the distribution of the intrinsic properties of LGRBs with the observed LGRB properties to estimate the unknown redshifts of individual BATSE LGRBs at the second stage of the inference process. The {\color{red}{red lines}} in the figure represent the unknown true intrinsic properties of individual BATSE LGRBs, whose knowledge is not essential in this redshift estimation workflow discussed above.}
        \label{fig:ToyProblem}
    \end{figure}

    Without loss of generality, suppose the observed properties of individual BATSE LGRBs are exactly known, with no measurement error, as illustrated by the individual blue-colored vertical lines in the bottom plot of Figure \ref{fig:ToyProblem}. The corresponding redshifts of these events, (represented by the black lines in the middle plot of Figure \ref{fig:ToyProblem}) are, however, unknown and we wish to estimate them. Although we have no knowledge of the joint population distribution of the intrinsic properties of LGRBs, illustrated by the red distribution in the top plot of Figure \ref{fig:ToyProblem}, there are strong arguments in favor of these properties being potentially well-described by a 4-dimensional multivariate log-normal distribution, $\mathcal{N}(\Mean,\CovMat)$ in the space of $\liso$, $\eiso$, $\epkz$, and $\durz$ \citep{shahmoradi2013multivariate,shahmoradi2015short}. Here, $\Mean$ and $\CovMat$ represent the mean and the covariance matrix of the multivariate log-normal distribution, respectively.\newpar

    We now reach the crucial step in the inference process: Despite the complete lack of information about the redshifts of BATSE LGRBs, we can use the existing prior knowledge about the overall cosmic redshift distribution of LGRBs to integrate over all possible redshifts for each observed LGRB in BATSE catalog to infer a range of plausible values for the intrinsic properties of the corresponding LGRB. These individually-computed probability density functions (PDFs) of the intrinsic properties can be then used to infer the unknown parameters $(\Mean,\CovMat)$ of the joint population distribution of the intrinsic properties of LGRBs (i.e., the multivariate log-normal distribution).\newpar

    Once $(\Mean,\CovMat)$  are constrained, we can use the inferred population distribution of the intrinsic LGRB properties together with the observed properties to estimate the redshifts of individual BATSE LGRBs, independently of each other. The estimated redshifts can be again used to further constrain $\mathcal{N}(\Mean,\CovMat)$ which will then result in even tighter estimates for the individual redshifts of BATSE LGRBs. This recursive progress can practically continue until convergence to a set of fixed individual redshift estimates occurs.\newpar

    At first glance, this simple semi-Bayesian mathematical approach may sound like magic and perhaps, too good to be true. Sometimes it is. However, as we later explain in \S\ref{sec:discussion}, it can also lead to reasonably-accurate results if some conditions regarding the problem and the observational dataset are satisfied.

\subsection{The Cosmic Rates of LGRBs}
\label{sec:methods:redshiftEstimation}

    More formally, we model the process of LGRB observation as a non-homogeneous Poisson process whose mean rate parameter is the `{\bf cen}sored' cosmic LGRB rate, $\mobs$. Representing each LGRB with,
    \begin{equation}
        \dinti = \{\dintilgrb,\zi\} ~,~1\leq i\leq 1366 ~, \label{eq:dint}
    \end{equation}

    \noindent where $\dintilgrb$ is defined by \eqref{eq:dintlgrb}, we compute the probability of occurrence of each BATSE LGRB event in the 5-dimensional attributes space, $\domaindint$, of $z$, $\liso$, $\eiso$, $\epkz$, $\durz$, as a function of the parameters, $\pobs$, of the observed LGRB rate model, $\mobs$,
    \begin{equation}
        \label{eq:modelGeneric}
        \pdfbig{ \dinti | \mobs , \pobs } \propto \mobs\big( \dinti , \pobs \big) ~,
    \end{equation}

    \noindent where $\mobs$ represents the BATSE-censored rate of LGRB occurrence in the universe (due to the BATSE detection efficiency limitations as detailed in \S\ref{sec:methods:modelingSampleIncompleteness}). This equation can be further expanded in terms of BATSE detection efficiency function, $\meff$, and the {\bf tru}e cosmic LGRB rate, $\mint$, as,
    \begin{eqnarray}
        \label{eq:modelobs}
        \frac{\diff\nobs}{\diff\dint}
        &=& \mobs\big( \dint , \pobs \big) ~, \nonumber \\
        &=& \meff\big( \dint , \peff \big) \times \mint\big( \dint , \pint \big) ~,
    \end{eqnarray}

    \noindent for a given set of input intrinsic LGRB attributes, $\dint$, with $\pobs=\{\peff,\pint\}$ as the set of the parameters of our models for the BATSE detection efficiency and the intrinsic cosmic LGRB rate, respectively. Assuming, no systematic evolution of LGRB characteristics with redshift, which is a plausible assumption supported by independent studies \citep[e.g.,][]{butler2010cosmic} (hereafter: \citetalias{butler2010cosmic}), the intrinsic LGRB rate itself can be written as,
    \begin{eqnarray}
        \label{eq:modelint}
        \frac{\diff\nint}{\diff\dint}
        &=& \mint \big( \dint , \pint \big) \nonumber \\
        &=& \mintlgrb \big( \dintlgrb , \pintlgrb \big) 
        \times \frac{\mz(z,\pz)\nicefrac{\diff V}{\diff z}}{(1+z)}~,~~~~
    \end{eqnarray}

    \noindent with $\pint=\{\pintlgrb,\pz\}$, where $\mintlgrb$ is a statistical model, with $\pintlgrb$ denoting its parameters, that describes the population distribution of LGRBs in the 4-dimensional attributes space of $\dintlgrb=[\liso,\eiso,\epkz,\durz]$, and the term $\mz(z,\pz)$ represents the comoving rate density model of LGRBs with the set of parameters $\pz$, while the factor $(1+z)$ in the denominator accounts for the cosmological time dilation. The comoving volume element per unit redshift, $\nicefrac{\diff V}{\diff z}$, is given by \citep[e.g.,][]{winberg1972gravitation, peebles1993principles},
    \begin{equation}
        \label{eq:dvdz}
        \frac{\diff V}{\diff z} = \frac{C}{H_0}\frac{4\pi {\ldis}^2(z)}{(1+z)^2\bigg[\Omega_M(1+z)^3+\Omega_\Lambda\bigg]^{1/2}} ~,
    \end{equation}

    \noindent with $\ldis$ standing for the luminosity distance as given in \eqref{eq:ldis}. If the three rate models, $(\mz,\meff,\mintlgrb)$, and their parameters were known a priori, one could readily compute the PDFs of the set of unknown redshifts of all BATSE LGRBs,
    \begin{equation}
        \label{eq:zset}
        \zset = \big\{ \zi: ~ 1\leq i\leq1366 \big\} ~,
    \end{equation}

    \noindent as,
    \begin{eqnarray}
        \label{eq:zPDF}
        \pi \big( \zset &|& \dsetobslgrb, \mobs, \pobs \big) \nonumber \\
        &\propto&   \int_{\domaindsetintlgrb} \mobs \big( \zset , \dsetintlgrb^* , \pobs \big) ~ \diff\dsetintlgrb^*,~~~~~~
    \end{eqnarray}

    \noindent where the integration is performed over all possible realizations, $\dsetintlgrb^*$, of the BATSE LGRB dataset given the measurement uncertainty models in \eqref{eq:dsetobslgrb}. For a range of possible parameter values, the redshift probabilities can be computed by marginalizing over the entire parameter space, $\domainpobs$, of the model,
    \begin{eqnarray}
        \label{eq:zMarginalPDF}
        \pi \big( \zset &|& \dsetobslgrb, \mobs \big) \nonumber \\
        &=& \int_\domainpobs \pdfbig{ \zset | \dsetobslgrb, \mobs, \pobs } \nonumber \\
        &\times& \pdfbig{ \pobs | \dsetobslgrb , \mobs } ~\diff\pobs ~.
    \end{eqnarray}

    The problem, however, is that neither the rate models nor their parameters are known a priori. Even more problematic is the circular dependency of the posterior PDFs of $\zset$ and $\pobs$ on each other,
    \begin{eqnarray}
        \label{eq:paraPostProb}
        \pi \big( \pobs &|& \dsetobslgrb , \mobs \big) \nonumber \\
        &=& \int_{\domain(\zset)} \pdfbig{ \pobs \big| \zset , \dsetobslgrb , \mobs } \nonumber \\
        &\times& \pi \big( \zset \big| \dsetobslgrb , \mobs \big) ~\diff\zset ~.
    \end{eqnarray}

     Therefore, we adopt the following methodology, which is reminiscent of the Empirical Bayes \citep[][]{robbins1985empirical} and Expectation-Maximization algorithms \citep[][]{dempster1977maximum}, to estimate the redshifts of BATSE LGRBs. First, we propose models for $(\mz,\meff,\mintlgrb)$, whose parameters have yet to be constrained by observational data. Given the three rate models, we can then proceed to constrain the free parameters of the observed cosmic LGRB rate, $\mobs$, based on BATSE LGRB data.\newpar

     The most appropriate fitting approach should take into account the observational uncertainties and any prior knowledge from independent sources. This can be achieved via the multilevel Bayesian methodology \citep[e.g.,][]{shahmoradi2017multilevel} by constructing the likelihood function and the posterior PDF of the parameters of the model, while taking into account the uncertainties in observational data \citep[e.g., Eqn. 61 in][]{shahmoradi2017multilevel},
    \begin{widetext}
        \begin{eqnarray}
            && \pi \big( \pobs | \dsetobslgrb , \mobs \big) \nonumber \\
            &=& \frac
            {
                \pdfbig{ \pobs \big| \mobs }
            }{
                \pdfbig{ \dsetobslgrb \big| \mobs }
            }
            \int_{\domain(\zset)} \int_{\domaindsetintlgrb} \pdfbig{ \dsetintlgrb^* \big| \dsetobslgrb, \zset , \mobs , \pobs } \pdfbig { \zset \big| \mobs , \pobs } ~ \diff\dsetintlgrb^* ~ \diff\zset \label{eq:paraPostGeneric} \\
            &\cpropto{i.i.d.}&
            \exp \bigg( -\int_\domaindint \mobs \big( \dintp , \pobs \big) \diff\dintp \bigg)
            \prod_{i=1}^{1366}
            \meff \big( \dobsilgrbmode , \peff \big)
            \int_\domaindint \mint \big( \dintp , \pint \big) \pdfbig{ \dintp ~|~ \dobsilgrb } \diff\dintp , ~~~~~~ \label{eq:paraPostPoisson} \\
            &\cequiv{no noise}&
            \exp \bigg( -\int_\domaindint \mobs \big( \dintp , \pobs \big) \diff\dintp \bigg)
            \prod_{i=1}^{1366}
            \meff \big( \dobsilgrbmode , \peff \big)
            \int_{z^*=0}^{z^*=+\infty} \mint \big( \dobsilgrbmode, z^*, \pint \big) \diff z^*, ~~~~~~ \label{eq:paraPostPoissonNoNois}
        \end{eqnarray}
    \end{widetext}

    \noindent where \eqref{eq:paraPostPoisson} holds under the assumption of independent and identical distribution (i.e., the i.i.d. property) of BATSE LGRBs, and the second integration within it is performed over all possible realizations of the truth for the $i$th observed BATSE LGRB, $\dobsilgrb$. Equation \eqref{eq:paraPostPoisson} can be further considerably simplified to \eqref{eq:paraPostPoissonNoNois} by assuming no measurement uncertainty in the observational data, except redshift ($z$) which is completely unknown for BATSE LGRBs.\newpar

    Once the posterior PDF of the model parameters is obtained, it can be plugged into \eqref{eq:zMarginalPDF} to constrain the redshift PDF of individual BATSE LGRBs at the second level of modeling.

\subsection{The LGRB Redshift Prior Knowledge}
\label{sec:methods:lgrbRateDensity}

    Our main assumption in this work is that the intrinsic comoving rate density of LGRBs closely traces the comoving Star Formation Rate (SFR) density \citep[e.g.,][]{madau2014cosmic, madau2017radiation, fermi2018gamma} (hereafter: \citetalias{madau2014cosmic, madau2017radiation, fermi2018gamma}, respectively) or a metallicity-corrected SFR density as prescribed by \citetalias{butler2010cosmic}. Consequently, regardless of the individually-unknown redshifts of BATSE LGRBs, the overall redshift distribution of all BATSE LGRBs together is enforced in our modeling to follow the cosmic SFR convolved with BATSE detection efficiency model, $\meff$, which is detailed in \S\ref{sec:methods:modelingSampleIncompleteness}. This assumption is essential for the success of our modeling approach, as any attempts to constrain the comoving rate density, $\mz(z)$, of LGRBs solely based on BATSE data leads to highly degenerate parameter space, $\domainpobs$, and parameter estimates for our model, $\mobs$.\newpar

    As for the choice of the LGRB rate density model, $\mz$, we have considered and simulated six different LGRB rate density scenarios, three of which have the generic continuous piecewise form,
    \begin{equation}
        \label{eq:mz}
        \mz(z) \propto
        \begin{cases}
            (1+z)^{\gamma_0} & z<z_0 \\
            (1+z)^{\gamma_1} & z_0<z<z_1 \\
            (1+z)^{\gamma_2} & z>z_1 ~, \\
        \end{cases}
    \end{equation}

    \noindent with parameters,
    \begin{eqnarray}
        \label{eq:pz}
        \pz
        &=& (z_0,z_1,\gamma_0,\gamma_1,\gamma_2) \nonumber \\
        &=&
        \begin{cases}
            (0.97,4.5,3.4,-0.3,-7.8)    & \text{(\citetalias{hopkins2006normalization})} \\
            (0.993,3.8,3.3,0.055,-4.46) & \text{(\citetalias{li2008star})} \\
            (0.97,4.00,3.14,1.36,-2.92) & \text{(\citetalias{butler2010cosmic})}
        \end{cases}
    \end{eqnarray}

    \noindent corresponding to the SFR density of \citet{hopkins2006normalization} (hereafter: \citetalias{hopkins2006normalization}), \citet{li2008star} (hereafter: \citetalias{li2008star}), and a bias-corrected redshift distribution of LGRBs derived from Swift data by \citetalias{butler2010cosmic}. The other three redshift scenarios follow the generic functional of Eqn. (15) in \citet{madau2014cosmic},
    \begin{equation}
        \label{eq:mzm}
        \mz(z) \propto \frac
        { (1+z)^{\alpha_1} }
        { (1+z)^{\alpha_2} + (1+z_0)^{\alpha_2} }
    \end{equation}

    \noindent with parameters,
    \begin{eqnarray}
        \label{eq:pzm}
        \pz
        &=& (z_0,\alpha_1,\alpha_2) \nonumber \\
        &=&
        \begin{cases}
            (1.90,2.70,5.60)    & \text{(\citetalias{madau2014cosmic})} \\
            (2.20,2.60,6.20)    & \text{(\citetalias{madau2017radiation})} \\
            (1.63,2.99,6.19)    & \text{(\citetalias{fermi2018gamma})}
        \end{cases}
    \end{eqnarray}

    For the sake of brevity, we will only present the results for four out of these six SFR density scenarios: \citetalias{hopkins2006normalization, butler2010cosmic, madau2017radiation, fermi2018gamma}.

\subsection{The LGRB Properties Rate Model: $\mintlgrb$}
\label{sec:methods:lgrbWorldModel}

    We model the joint 4-dimensional distribution of $\dintlgrb$ with a multivariate log-normal distribution, $\mintlgrb\equiv\mathcal{LN}$, whose parameters (i.e., the mean vector and the covariance matrix), $\pintlgrb=\{\bs\mu,\bs\Sigma\}$, will have to be constrained by data. The justification for the choice of a multivariate log-normal as the underlying intrinsic population distribution of LGRBs is multi-folded. First, the observed joint distribution of BATSE LGRB properties highly resembles a log-normal shape that is censored close to the detection threshold of BATSE. Second, unlike power-law distribution which has traditionally been the default choice of model for the luminosity function of LGRBs, log-normal models provide natural upper and lower bounds on the total energy budget and luminosity of LGRBs, eliminating the need for setting artificial sharp bounds on the distributions to properly normalize them. Third, log-normal along with Gaussian distribution, are among the most naturally-occurring statistical distributions in nature, whose generalizations to multi-dimensions are also well studied and understood. This is a highly desired property especially for our work, given the overall mathematical and computational complexity of the model proposed and developed here.

\subsection{The BATSE Trigger Efficiency: $\meff$}
\label{sec:methods:modelingSampleIncompleteness}

    Compared to Fermi Gamma-Ray Burst Monitor \citep[][]{meegan2009fermi} and Neil Gehrels Swift Observatory \citep[][]{gehrels2004swift, lien2016third}, BATSE had a relatively simple triggering algorithm. The BATSE detection efficiency and algorithm has been already extensively studied by the BATSE team as well as independent authors \citep[e.g.,][]{pendleton1998batse, pendleton1995detector, hakkila2003sample} \citep[c.f.,][for further discussion and references]{shahmoradi2010hardness, shahmoradi2011possible, shahmoradi2013multivariate, shahmoradi2015short}. However, simple implementation and usage of the known BATSE trigger threshold for modeling the BATSE catalog's sample incompleteness can lead to systematic biases in the inferred quantities of interest. BATSE triggered on $2702$ GRBs, out of which only $2145$, or approximately $79\%$, have been consistently analyzed and reported in the current BATSE catalog, with the remaining $21\%$ either having a low accumulation of count rates or missing a full spectral/temporal coverage \citep[][]{goldstein2013batse}. Thus, the extent of sample incompleteness in the BATSE catalog is likely not fully and accurately represented by the BATSE triggering algorithm alone.\newpar

\begin{figure}
        \centering
        \includegraphics[width=0.45\textwidth]{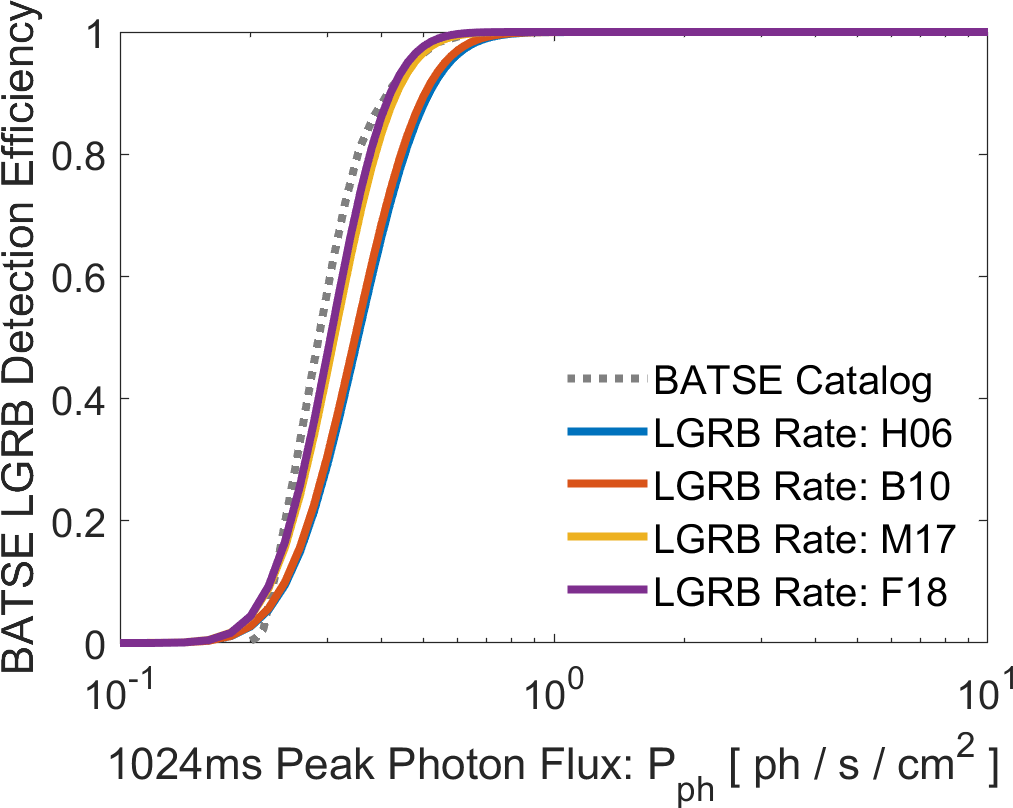}
        \caption{A comparison of the BATSE 4B Catalog nominal LGRB detection efficiency as a function of the 1024 [ms] peak photon flux, $\pph$, with the predicted detection efficiencies in this work based on the four different LGRB rate models considered: \citetalias{hopkins2006normalization}, \citetalias{butler2010cosmic}, \citetalias{madau2017radiation}, \citetalias{fermi2018gamma}. The peak photon flux, $\pph$, is measured in the BATSE energy window $50-300$ [keV].}
        \label{fig:BatseDetEff}
    \end{figure}

    BATSE LADs generally triggered on a GRB if the number of photons per 1024 [ms] arriving at the detectors in $50-300$ [keV] energy window, $\pph$, reached a certain threshold in units of the background photon count fluctuations, $\sigma$. This threshold was typically set to $5.5\sigma$ during much of BATSE's operational lifetime. However, the naturally-occurring fluctuations in the average background photon counts effectively lead to a monotonically increasing BATSE detection efficiency as a function of $\pph$, instead of a sharp cutoff on the observed $\pph$ distribution of LGRBs. Therefore, we model the effects of BATSE detection efficiency and sample incompleteness more accurately by an Error function,
    \begin{eqnarray}
        \label{eq:methods:redshiftEstimation:efficiency}
        \pi\big(\mathrm{detection}&|&\threshM,\threshS,\pph\big) \nonumber \\
        &=& \frac{1}{2} + \frac{1}{2} \times \mathrm{erf}\bigg(\frac{\logten\pph-\threshM}{\threshS\sqrt{2}}\bigg) ~,
    \end{eqnarray}

    \noindent that, for a given set of Error function's parameters ($\peff=\{\threshM,\threshS\}$) yields the probability of the detection of an LGRB with $\pph$ photon counts per second. Due to the unknown effects of sample incompleteness on BATSE data, we leave these two threshold parameters free solely to be constrained by the observational data. Figure \ref{fig:BatseDetEff} compares the resulting best-fit detection efficiency functions for BATSE under different SFR scenarios with the BATSE nominal detection efficiency for LGRBs\footnote{Available at: \url{https://gammaray.nsstc.nasa.gov/batse/grb/catalog/4b/4br_efficiency.html}} \citep[see also][]{pendleton1995detector, pendleton1998batse, paciesas1999fourth, hakkila2003sample, shahmoradi2013multivariate, shahmoradi2015short}.\newpar

    The current BATSE catalog already provides estimates of $\pph$ for all 1366 LGRBs in our analysis. The connection between $\pph$ and the bolometric 1024 [ms] peak flux, $\pbol$, which is used in our modeling, is provided by fitting all LGRB spectra with a smoothly-broken power-law,
    \begin{equation}
        \label{eq:Band}
        \Phi (E) \propto
        \begin{cases}
            E^{\alpha}~ \operatorname{e}^{\big(-\frac{(1+z)(2+\alpha)E}{\epkz}\big)} & \text{if $E\le\big(\frac{\epkz}{1+z}\big)\big(\frac{\alpha-\beta}{2+\alpha}\big)$,} \\
            E^{\beta} & \text{if otherwise.}
        \end{cases}
    \end{equation}

    \noindent known as the Band model \citep[][]{band1993batse} to infer their spectral normalization constants while fixing the low- and high-energy photon indices of the Band model to the corresponding population averages $(\alpha, \beta) = (-1.1,-2.3)$ and fixing the observed spectral peak energies, $\epk$, of individual bursts to the corresponding best-fit values from \citet{shahmoradi2010hardness}. Such an approximation is reasonable given the typically large uncertainties that exist in the spectral and temporal properties of GRBs \citep[e.g.,][]{butler2010cosmic, shahmoradi2013multivariate} and the relatively large variance of the population distribution of $\pbol$ in our sample.

\section{Results}
\label{sec:results}

    \begin{figure*}[tphb]
        \centering
        \begin{tabular}{cccc}
            \includegraphics[width=0.233\textwidth]{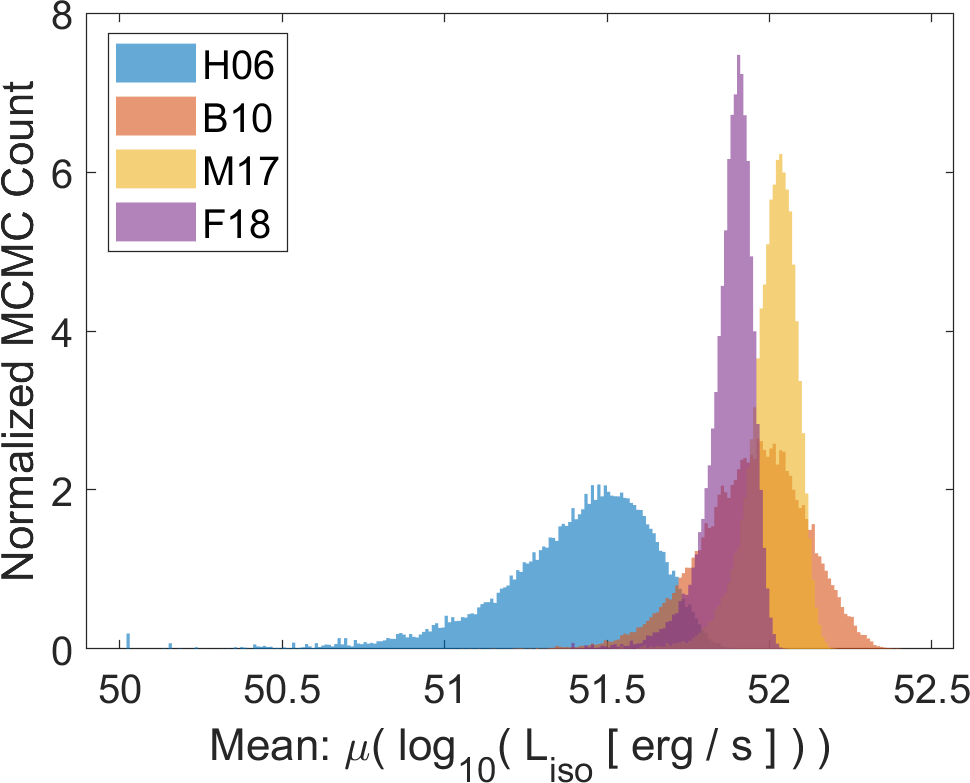} &
            \includegraphics[width=0.233\textwidth]{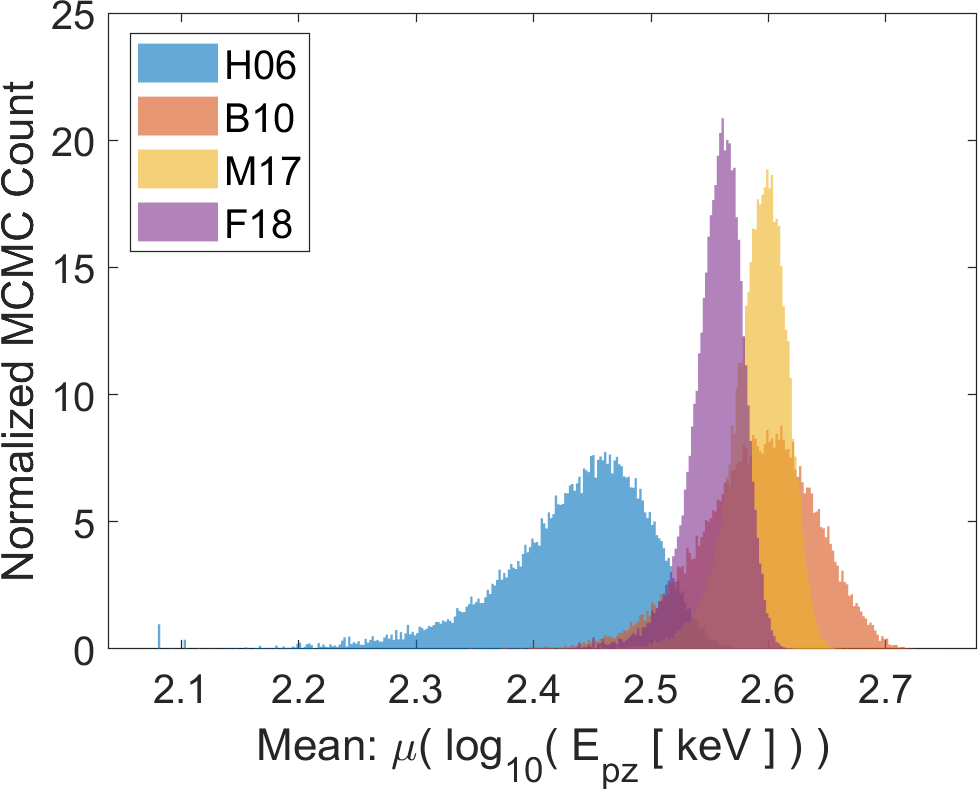} &
            \includegraphics[width=0.233\textwidth]{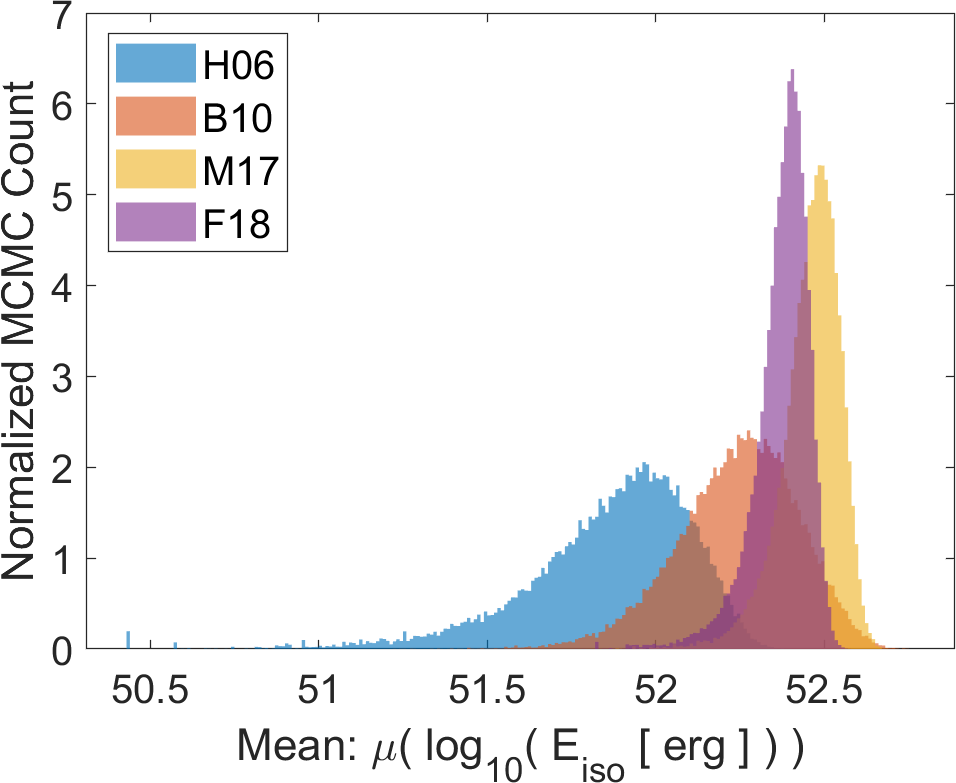} &
            \includegraphics[width=0.233\textwidth]{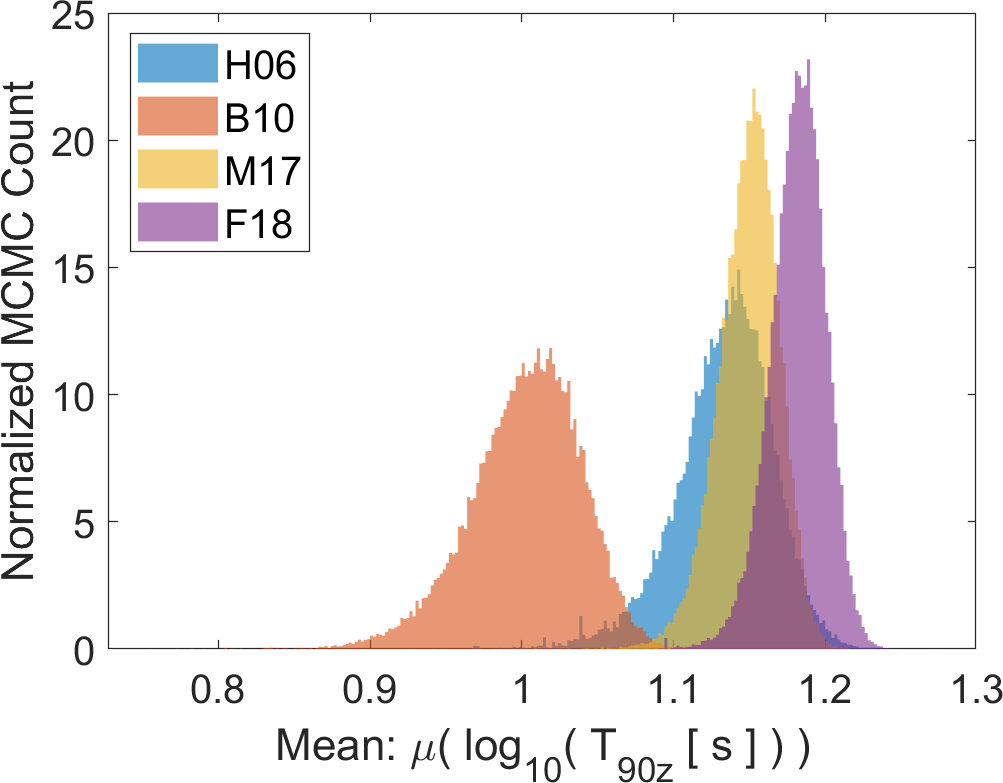} \\
            \includegraphics[width=0.233\textwidth]{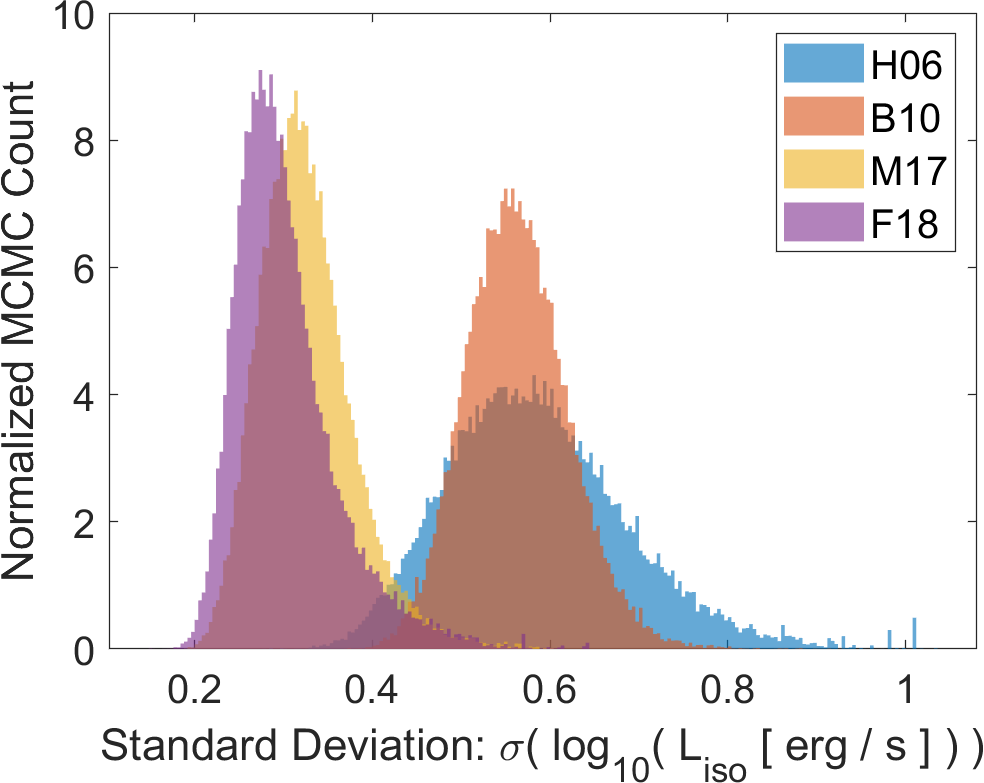} &
            \includegraphics[width=0.233\textwidth]{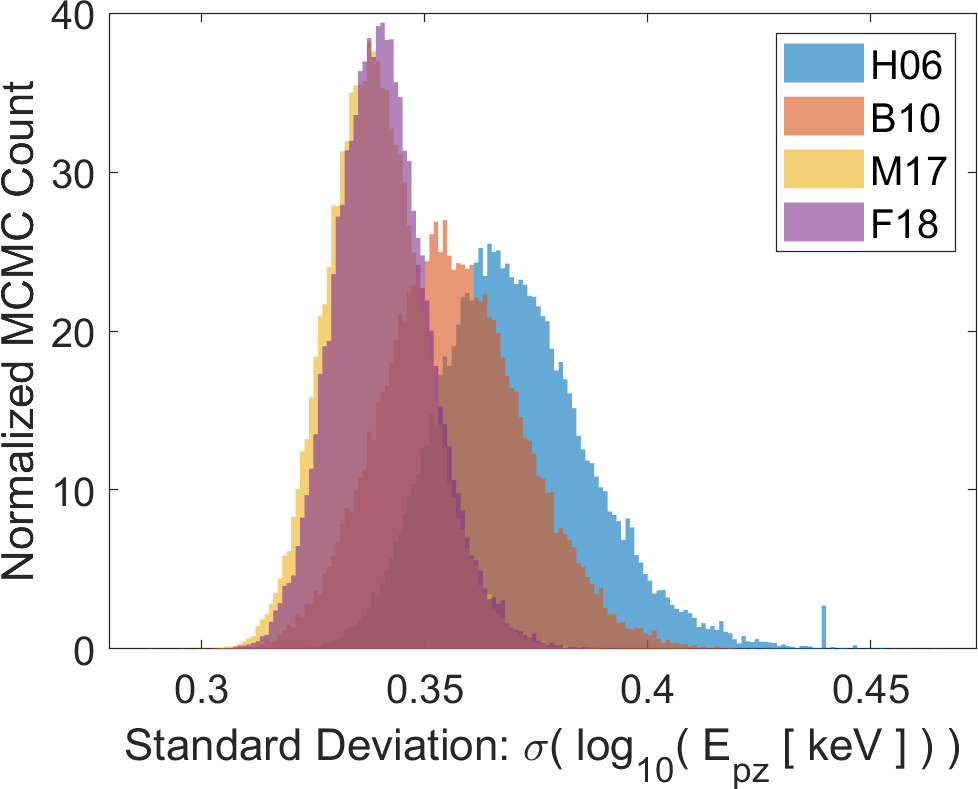} &
            \includegraphics[width=0.233\textwidth]{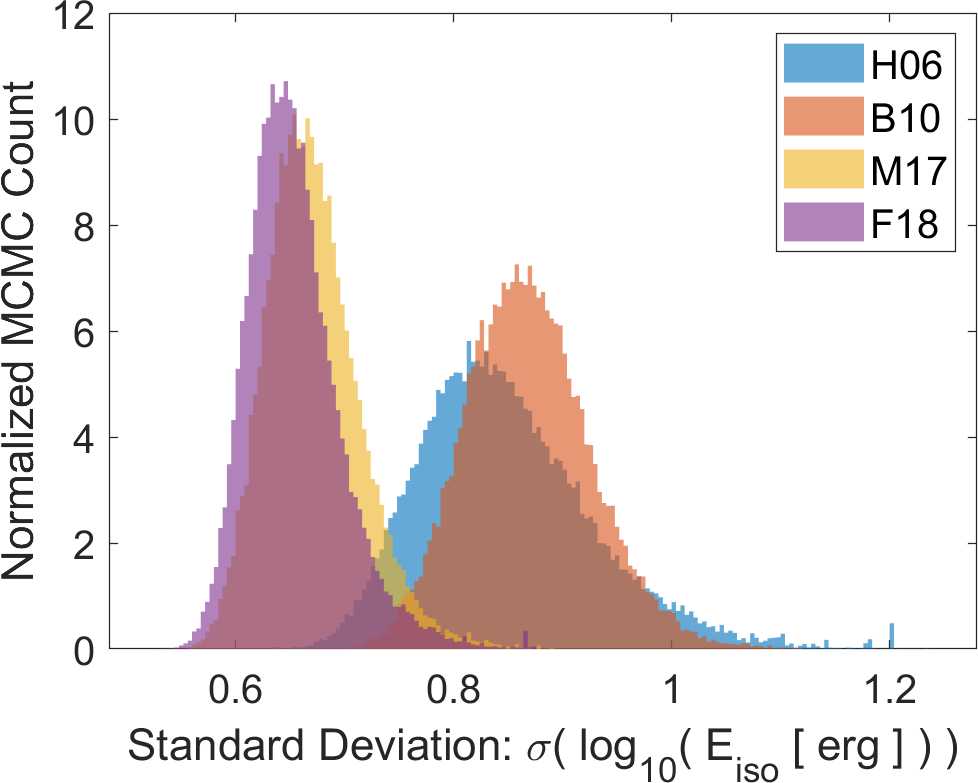} &
            \includegraphics[width=0.233\textwidth]{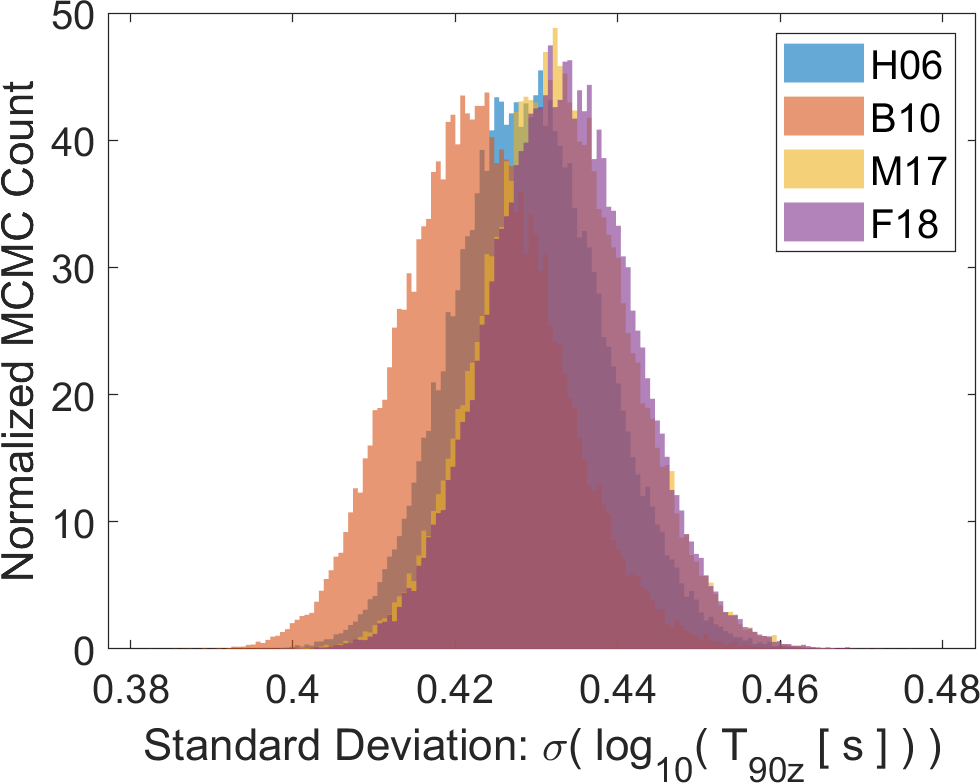} \\
            \includegraphics[width=0.233\textwidth]{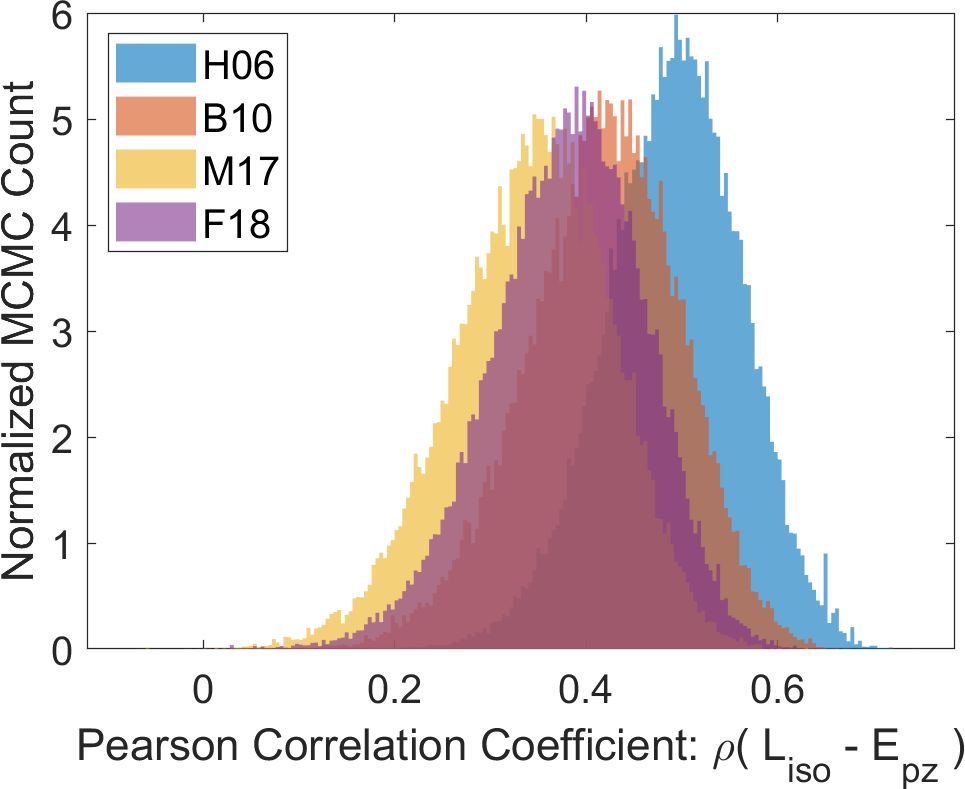} &
            \includegraphics[width=0.233\textwidth]{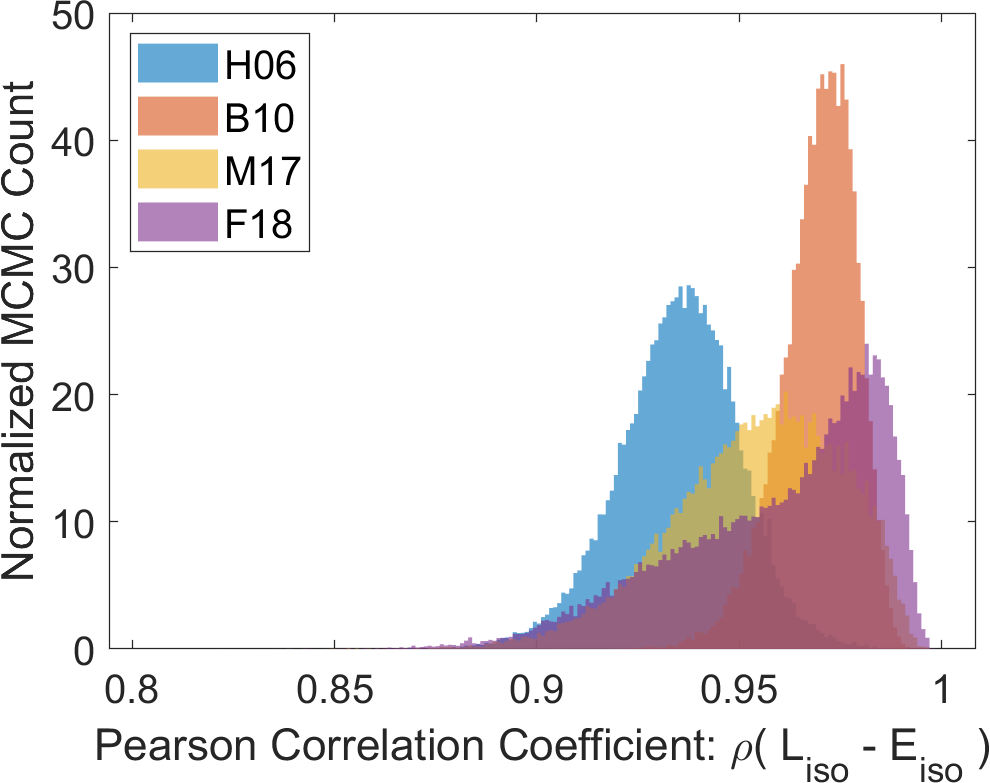} &
            \includegraphics[width=0.233\textwidth]{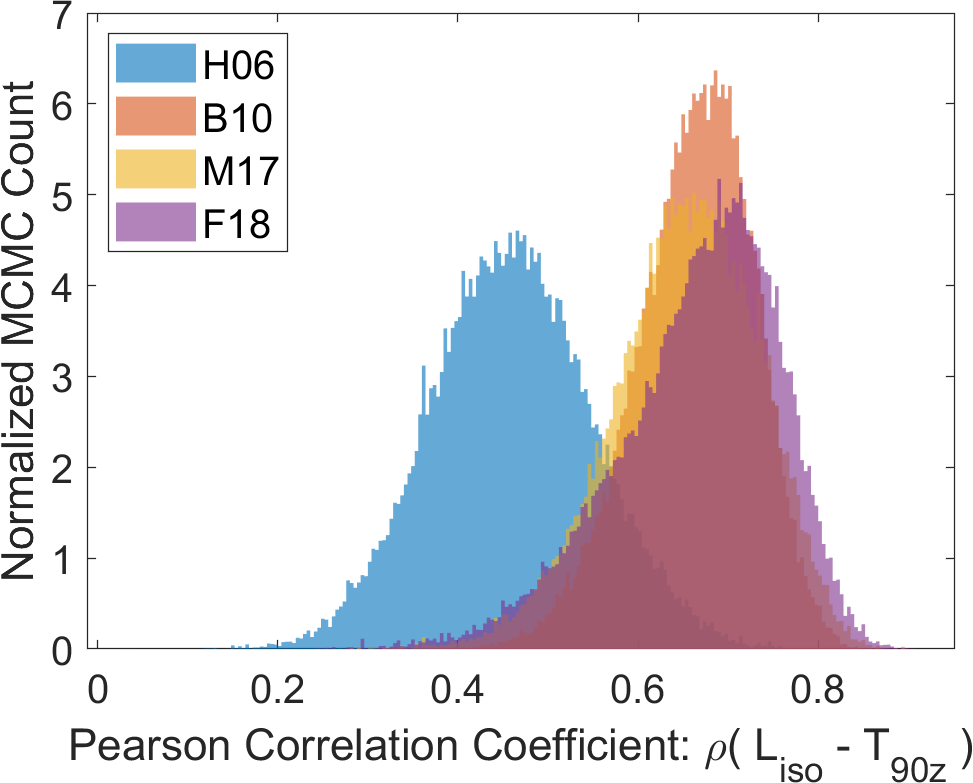} &
            \includegraphics[width=0.233\textwidth]{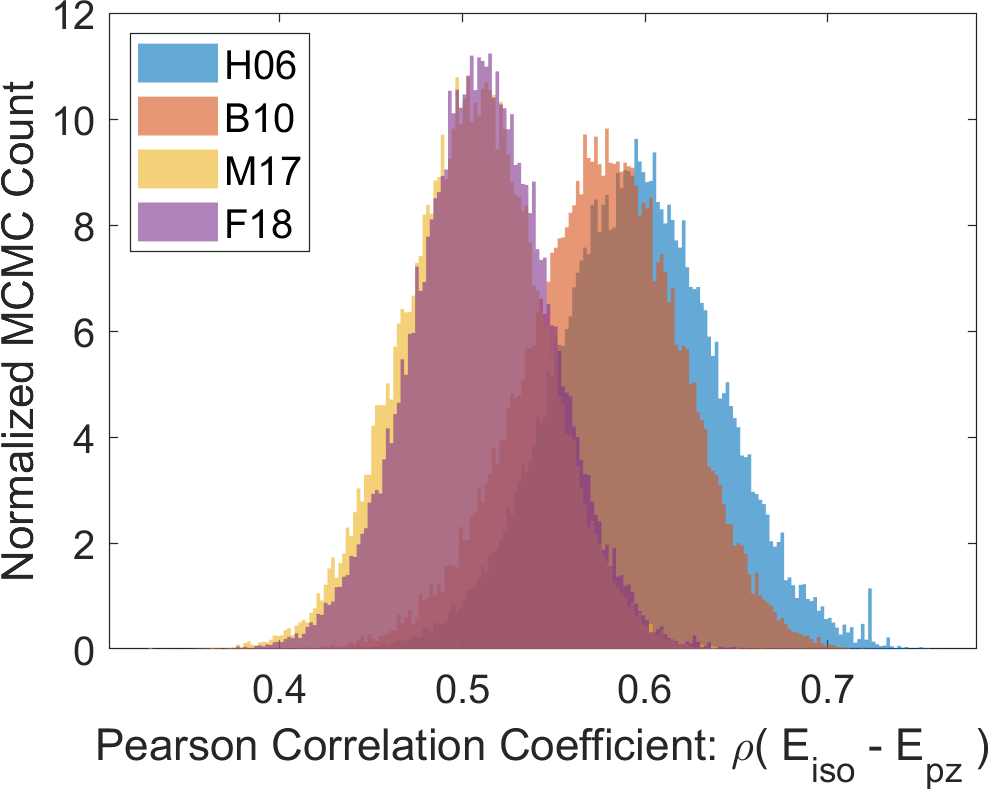} \\
            \includegraphics[width=0.233\textwidth]{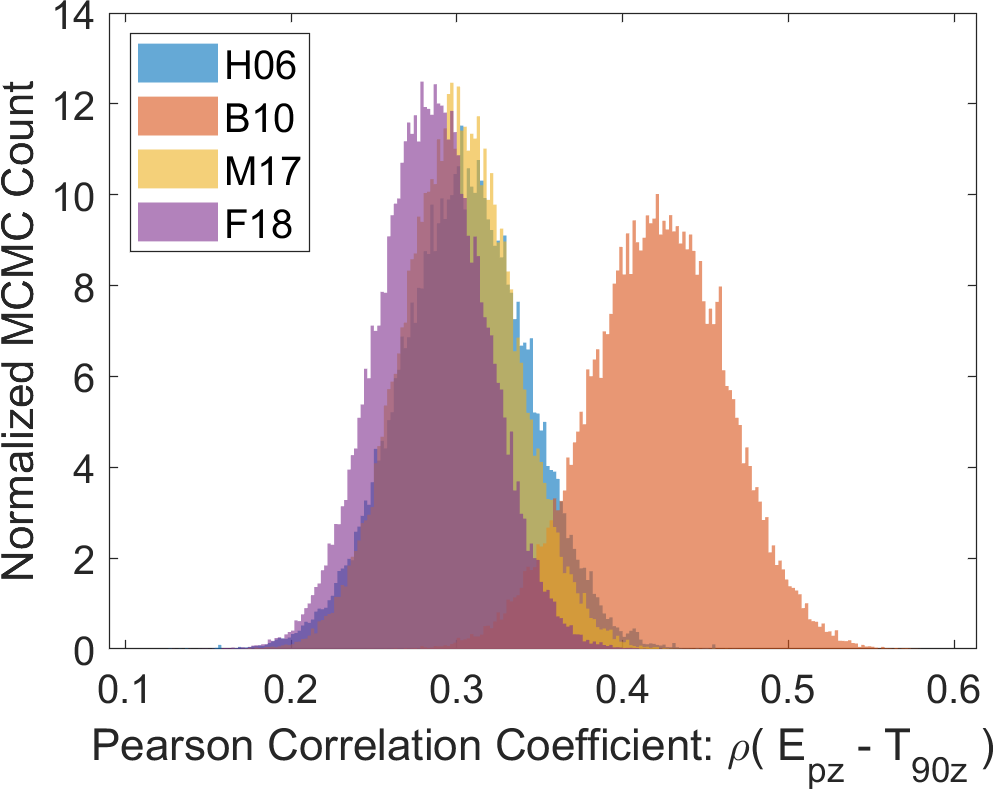} &
            \includegraphics[width=0.233\textwidth]{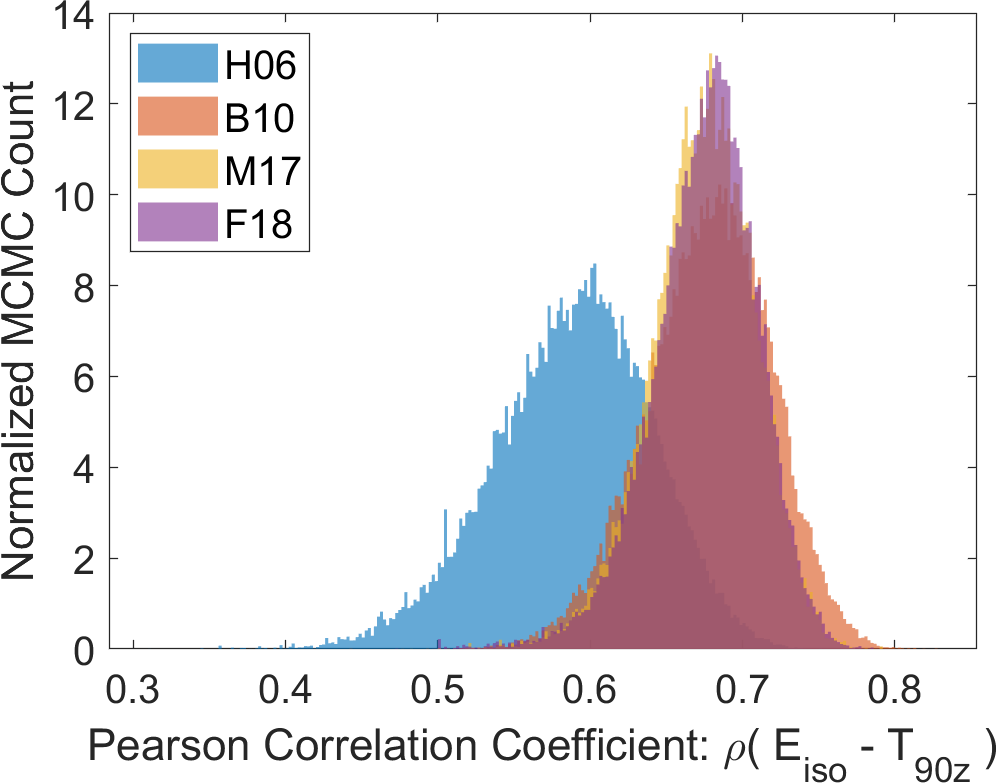} &
            \includegraphics[width=0.233\textwidth]{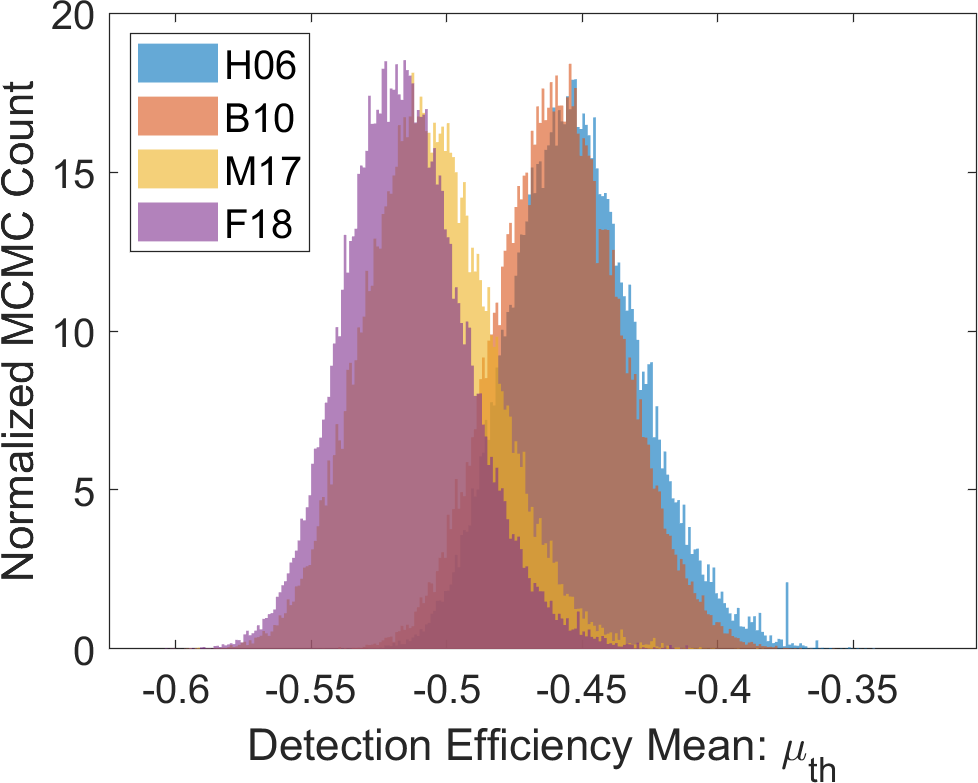} &
            \includegraphics[width=0.233\textwidth]{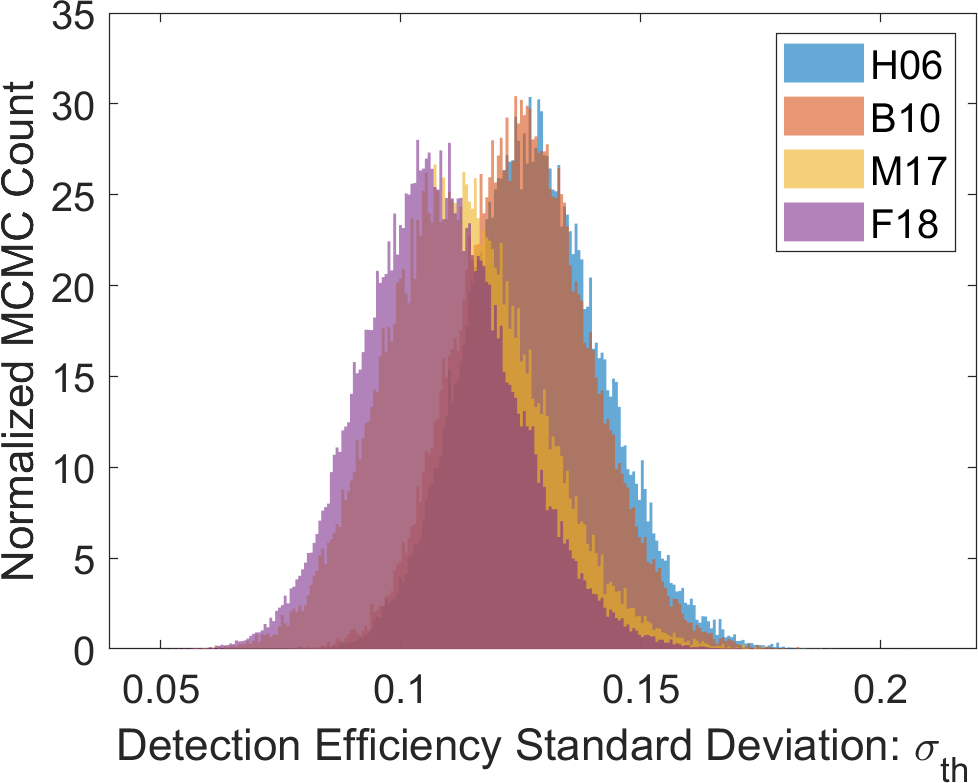} \\
        \end{tabular}
        \caption{The marginal posterior distributions of the 16 parameters of the LGRB world model, for the four redshift distributions considered in this work: \citetalias{hopkins2006normalization}, \citetalias{butler2010cosmic}, \citetalias{madau2017radiation}, \citetalias{fermi2018gamma}, as given by \eqref{eq:mz}, \eqref{eq:pz}, \eqref{eq:mzm}, and \eqref{eq:pzm}.\label{fig:paraPostMarginal}}
    \end{figure*}

    \begin{table*}[t]
    \begin{center}
        \vspace{5mm}
        \caption{Mean best-fit parameters of the LGRB World Model, \eqref{eq:modelGeneric}, for the four redshift distribution scenarios considered.\label{tab:paraPostStat}}
        \begin{tabular}{c c c c c }
            \hline
            \hline
            Parameter & \citetalias{hopkins2006normalization} & \citetalias{butler2010cosmic} & \citetalias{madau2017radiation} & \citetalias{fermi2018gamma} \\
            \hline
            \multicolumn{5}{c}{Location Parameters($\Mean$)} \\
            \hline
            $\mu_{\logten(\liso)}$ & $51.40\pm0.24$ & $51.95\pm0.16$ & $52.01\pm0.08$ & $51.88\pm0.08$ \\
            $\mu_{\logten(\epkz)}$ & $2.43\pm0.06$ & $2.59\pm0.05$ & $2.59\pm0.02$ & $2.56\pm0.02$ \\
            $\mu_{\logten(\eiso)}$ & $51.86\pm0.25$ & $52.24\pm0.18$ & $52.46\pm0.09$ & $52.37\pm0.08$ \\
            $\mu_{\logten(\durz)}$ & $1.13\pm0.03$ & $1.00\pm0.04$ & $1.15\pm0.02$ & $1.18\pm0.02$ \\
            \hline
            \multicolumn{5}{c}{Scale Parameters (diagonal elements of $\CovMat$)} \\
            \hline
            $\sigma_{\logten(\liso)}$ & $0.59\pm0.10$ & $0.56\pm0.06$ & $0.33\pm0.05$ & $0.30\pm0.06$ \\
            $\sigma_{\logten(\epkz)}$ & $0.37\pm0.02$ & $0.36\pm0.02$ & $0.34\pm0.01$ & $0.34\pm0.01$ \\
            $\sigma_{\logten(\eiso)}$ & $0.85\pm0.08$ & $0.87\pm0.06$ & $0.67\pm0.04$ & $0.65\pm0.04$ \\
            $\sigma_{\logten(\durz)}$ & $0.43\pm0.01$ & $0.42\pm0.01$ & $0.43\pm0.01$ & $0.43\pm0.01$ \\
            \hline
            \multicolumn{5}{c}{Correlation Coefficients (non-diagonal elements of $\CovMat$)} \\
            \hline
            $\rho_{\liso-\epkz}$ & $0.49\pm0.07$ & $0.42\pm0.08$ & $0.34\pm0.08$ & $0.38\pm0.08$ \\
            $\rho_{\liso-\eiso}$ & $0.93\pm0.01$ & $0.97\pm0.01$ & $0.95\pm0.02$ & $0.96\pm0.03$ \\
            $\rho_{\liso-\durz}$ & $0.46\pm0.09$ & $0.66\pm0.06$ & $0.65\pm0.08$ & $0.67\pm0.09$ \\
            $\rho_{\epkz-\eiso}$ & $0.60\pm0.04$ & $0.58\pm0.04$ & $0.51\pm0.04$ & $0.51\pm0.04$ \\
            $\rho_{\epkz-\durz}$ & $0.30\pm0.04$ & $0.42\pm0.04$ & $0.30\pm0.03$ & $0.28\pm0.03$ \\
            $\rho_{\eiso-\durz}$ & $0.59\pm0.05$ & $0.68\pm0.04$ & $0.67\pm0.03$ & $0.68\pm0.03$ \\
            \hline
            \multicolumn{5}{c}{BATSE Detection Efficiency (Sample Incompleteness)} \\
            \hline
            $\mu_{th}$ & $-0.45\pm0.02$ & $-0.46\pm0.02$ & $-0.51\pm0.02$ & $-0.51\pm0.02$ \\
            $\sigma_{th}$ & $0.13\pm0.01$ & $0.13\pm0.01$ & $0.11\pm0.02$ & $0.11\pm0.02$ \\
            \hline
            \hline
        \end{tabular}
    \end{center}
\end{table*}

    We proceed by first fitting the proposed censored cosmic LGRB rate model, $\mobs$ to 1366 BATSE LGRB data under the six LGRB redshift distribution scenarios prescribed by \eqref{eq:mz}--\eqref{eq:pz} and \eqref{eq:mzm}--\eqref{eq:pzm}. For each LGRB rate density scenario, the posterior PDF of parameters in \eqref{eq:paraPostPoisson} is explored by a {\bf Para}llel {\bf D}elayed-{\bf R}ejection {\bf A}daptive {\bf M}etropolis-Hastings Markov Chain Monte Carlo algorithm (the ParaDRAM algorithm) that we have developed for such sampling tasks as part of a larger Monte Carlo simulation package named ParaMonte available in C/C++/Fortran/MATLAB/Python programming languages\footnote{Available at: \url{https://github.com/cdslaborg/paramonte}} \citep[e.g.,][]{shahmoradi2019paramonte, kumbhare2020parallel, shahmoradi2020paramonte}.\newpar

    The computations for each SFR scenario are performed on 96 processors in parallel on two Skylake compute-nodes of the Stampede2 supercomputer at Texas Advanced Computing Center. We ran extensive tests to ensure a high-level of accuracy of the high-dimensional numerical integrations involved in the derivation of the posterior distribution of the parameters of the censored cosmic rate model for LGRBs as given in \eqref{eq:paraPostGeneric}. For the sake of brevity, the resulting best-fit parameters corresponding to only four out of these six SFR densities are tabulated in Table \ref{tab:paraPostStat}, and the marginal distributions of their parameters are compared with each other in Figure \ref{fig:paraPostMarginal}.\newpar

    \begin{figure*}[t!]
        \centering
        \begin{tabular}{ccc}
            \includegraphics[width=0.316\textwidth]{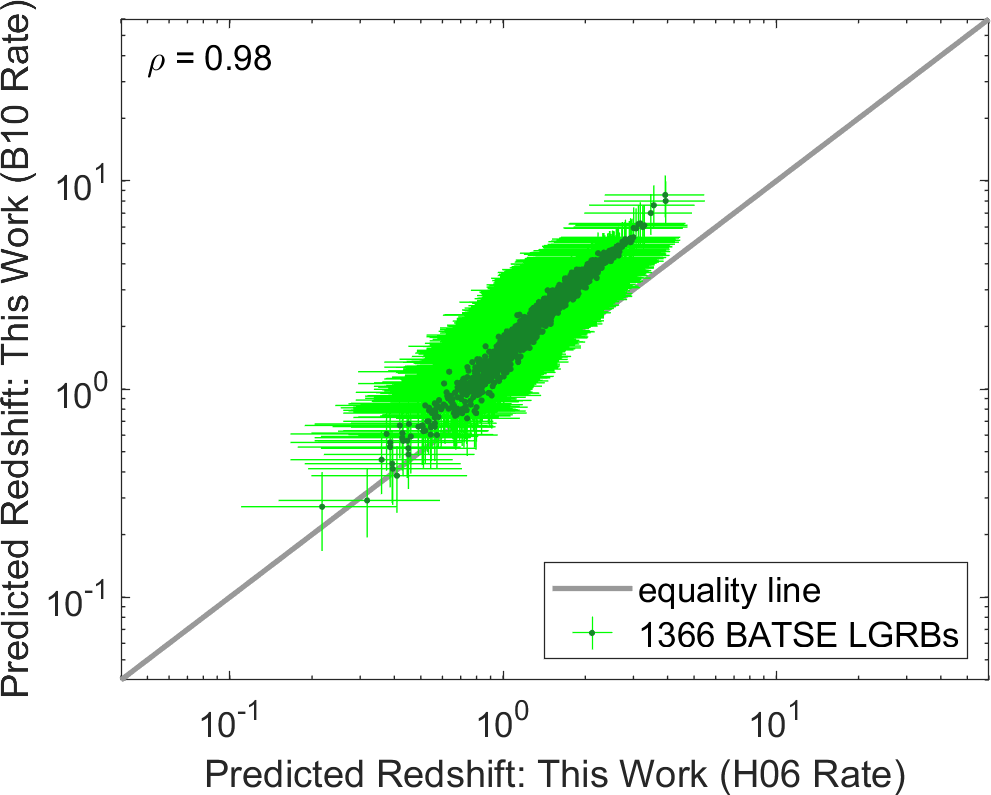} &
            \includegraphics[width=0.316\textwidth]{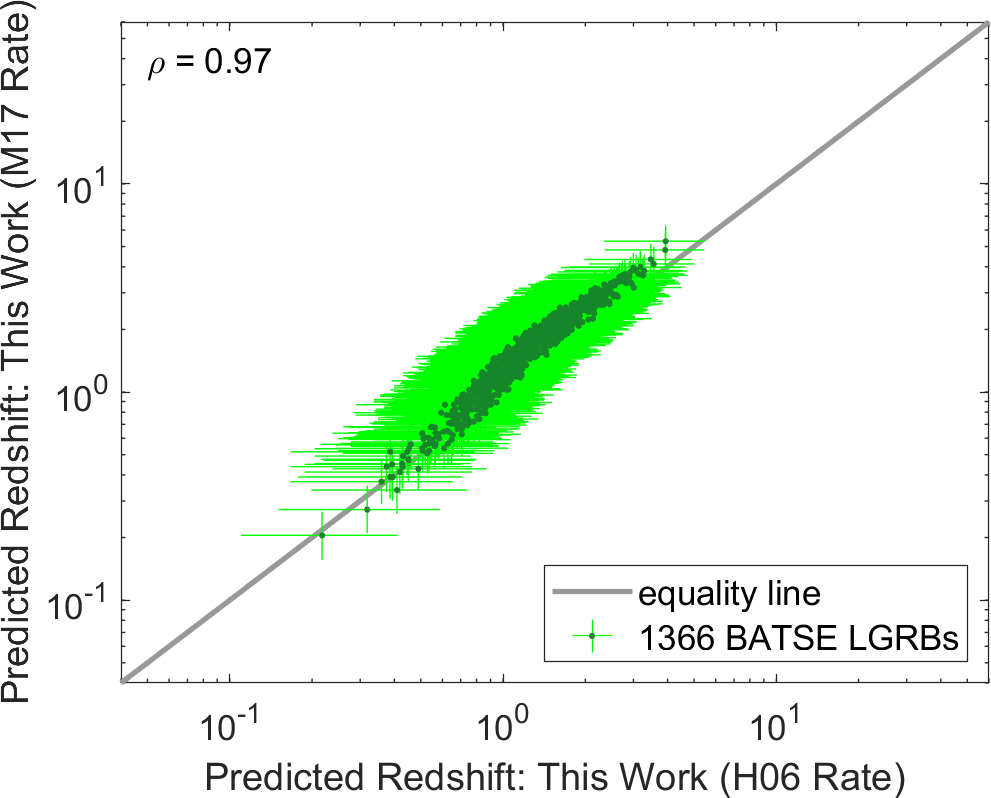} &
            \includegraphics[width=0.316\textwidth]{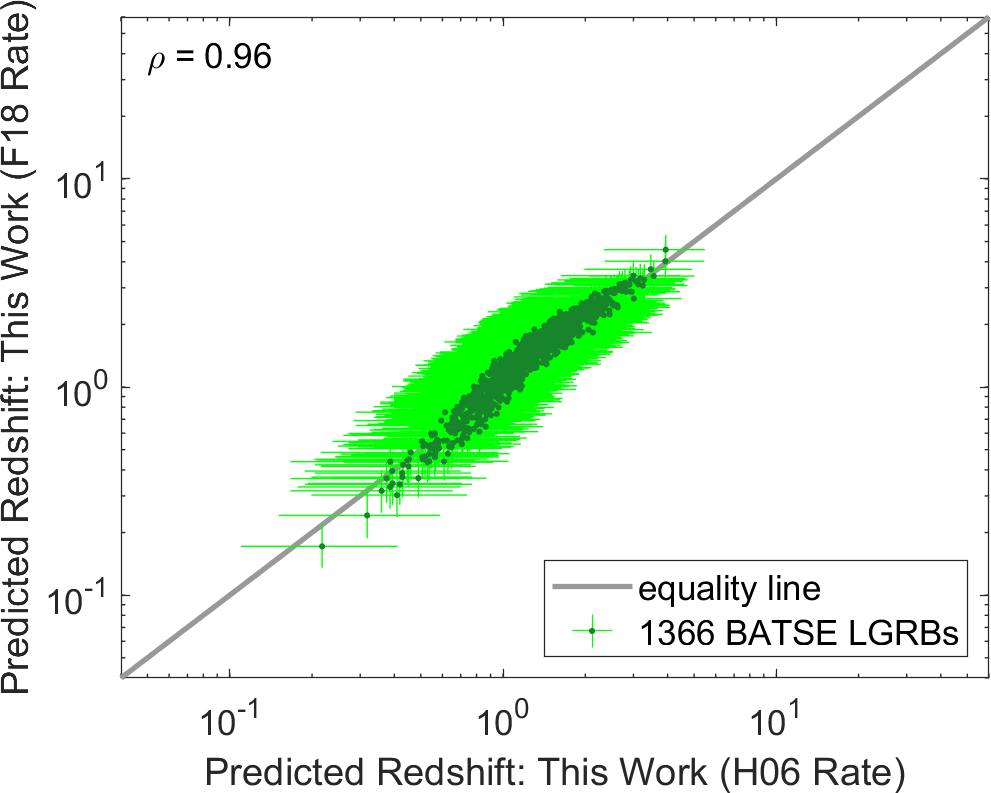} \\
            \includegraphics[width=0.316\textwidth]{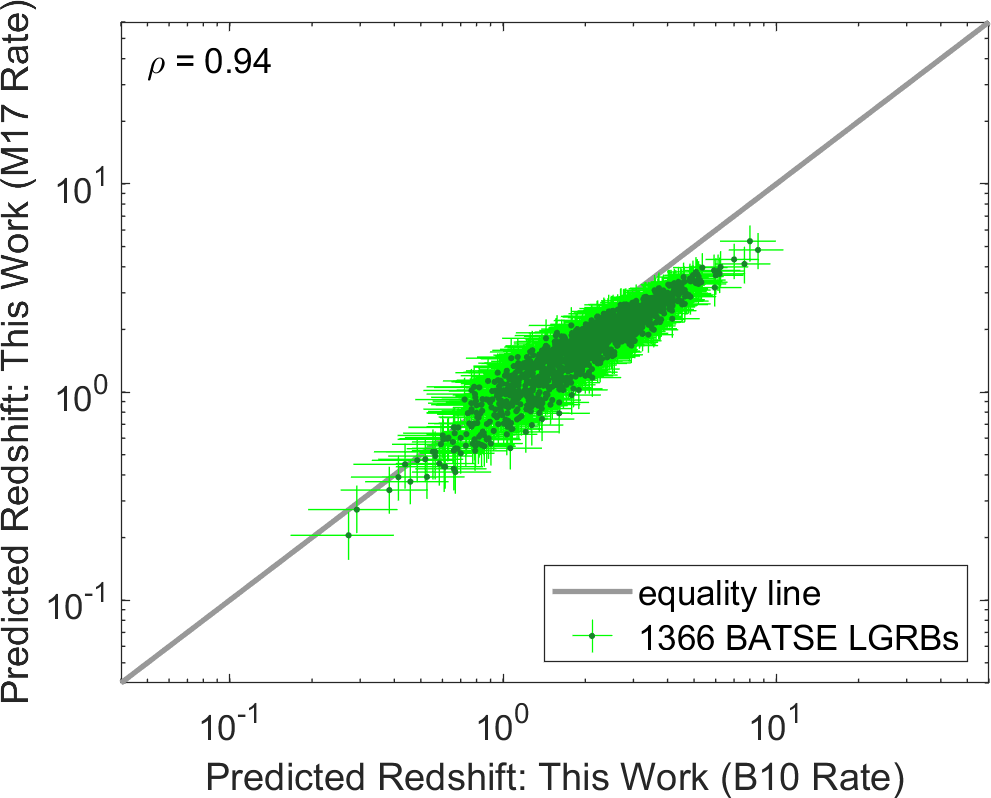} &
            \includegraphics[width=0.316\textwidth]{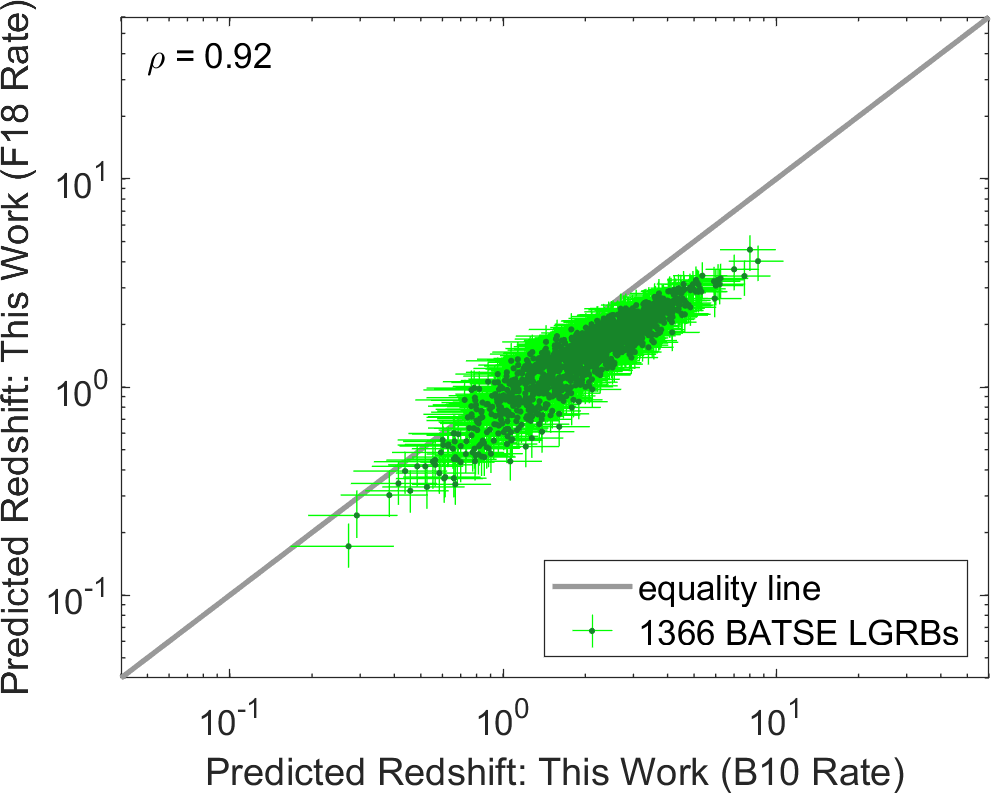} &
            \includegraphics[width=0.316\textwidth]{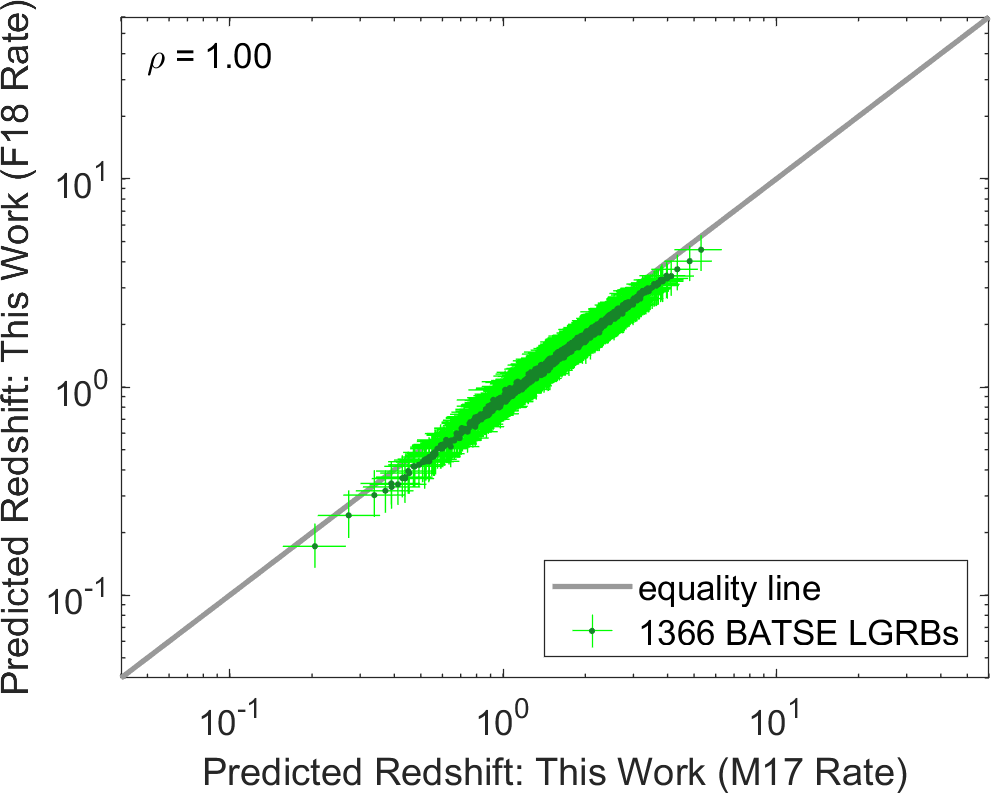} \\
        \end{tabular}
        \caption{A comparison of the expected redshifts of 1366 BATSE LGRBs for the four different cosmic LGRB rate density assumptions \eqref{eq:mz} with each other. On each plot, the Spearman's correlation coefficient of the two sets of expected redshifts is also reported. The error bars represent the $90\%$ prediction intervals of each individual redshift.
        \label{fig:expectedRedshift}}
    \end{figure*}

    Once the parameters of the censored cosmic rate model \eqref{eq:modelobs} are constrained, we use the calibrated model at the second level of our analysis to further constrain the PDFs of the unknown redshifts of individual BATSE LGRBs according to \eqref{eq:zMarginalPDF}. Similar to the Empirical Bayes methodology, this iterative process can continue until convergence to a specific set of redshift PDFs occurs. However, given the computational complexity and the expense of each iteration, we stop after obtaining the first round of estimates. This is also a common practice in the Empirical Bayes modeling. To further reduce the computational complexities of the simulations, we also drop the measurement uncertainties described by \eqref{eq:dobsilgrb} in all observational data from both levels of the analysis in \eqref{eq:paraPostPoisson} and \eqref{eq:zMarginalPDF}. To further reduce the computational expense of the inference, we also approximate the numerical integration in the definition of the luminosity distance in \eqref{eq:ldis} by the analytical expressions of \citet{wickramasinghe2010analytical}. All routines were implemented in the Fortran programming language and comply with the latest Fortran Standard published in 2018 \citep[e.g.,][]{metcalf2011modern, metcalf2018modern, reid2018new}\footnote{Available at:\\\url{https://github.com/cdslaborg/BatseLgrbRedshiftCatalog}}. The complete set of 1366 BATSE LGRBs observational data is also publicly available for download at the project's permanent repository on GitHub.\newpar

    The mean redshifts together with the $90\%$ prediction intervals for the four LGRB rate density scenarios are also reported in Table \ref{tab:redshiftCatalog}.
    We find that, on average, the redshifts of individual BATSE LGRBs can be constrained to within 50\% uncertainty ranges of 0.55, 0.46, 0.25, and 0.19 corresponding to the four LGRB rate densities of \citetalias{hopkins2006normalization}, \citetalias{butler2010cosmic}, \citetalias{madau2017radiation}, and \citetalias{fermi2018gamma}, respectively. At 90\% confidence level, the prediction intervals expand to wider uncertainty ranges of 1.40, 1.15, 0.71, and 0.60, respectively.

\section{discussion}
    \label{sec:discussion}


    In this work, we proposed a semi-Bayesian methodology to infer the unknown redshifts of 1366 BATSE catalog LGRBs. Towards this, first, we segregated the two populations of BATSE LGRBs and SGRBs using the fuzzy C-means classification method based on the observed duration and spectral peak energies of 1966 BATSE GRBs with available spectral and temporal information. We then modeled the process of LGRB detection as a non-homogeneous spatiotemporal Poisson process, whose rate parameter is modeled by a multivariate log-normal distribution as a function of the four main LGRB intrinsic attributes: the 1024 [ms] isotropic peak luminosity ($\liso$), the total isotropic emission ($\eiso$), the intrinsic spectral peak energy ($\epkz$), and the intrinsic duration ($\durz$). To calibrate the parameters of the rate model, we made a fundamental assumption that LGRBs trace the Cosmic Star Formation Rate (SFR), or a metallicity-corrected SFR. For each of the individual LGRB rate densities considered in this work: \citetalias{hopkins2006normalization}, \citetalias{butler2010cosmic}, \citetalias{madau2017radiation}, \citetalias{fermi2018gamma}, we then used the resulting posterior probability densities of the model parameters to compute the probability density functions of the redshifts of individual BATSE LGRBs.\newpar

    As illustrated in Figure \ref{fig:expectedRedshift}, we find that the individual redshift estimates of BATSE LGRBs for the LGRB rate density assumptions are broadly consistent with each other. There are, however, systematic differences between the estimates, the most significant of which is the difference between the expected redshifts based on \citetalias{butler2010cosmic} and the other rate densities. The difference can be explained by the larger rates of LGRBs that the model of \citetalias{butler2010cosmic} implies at higher redshifts compared to the other rate scenarios which exactly model the SFR in the universe. This, in effect, shifts the expected redshifts of BATSE LGRBs systematically toward larger values.\newpar

    As we explained in \S\ref{sec:methods:bayesianToyProblem}, the underlying logic of our approach to resolving the individual redshifts of BATSE LGRBs can be qualitatively understood by reconsidering the set of equations in \eqref{eq:obsIntMap}. Taking the logarithm of both sides of all equations, we obtain a set of linear maps from the rest-frame to the observer-frame properties of LGRBs \eqref{eq:obsIntMapLinearized}. Therefore, the distributions of the four observer-frame LGRB properties result from the convolution of the distributions of the corresponding rest-frame LGRB properties with the distribution of terms that are exactly determined by redshift (i.e., the logarithm of the luminosity distance, $\logten(\ldis)$ and the term $\logten(z+1)$).\newpar

    \begin{figure*}[t!]
        \centering
        \begin{tabular}{cccc}
            \includegraphics[width=0.235\textwidth]{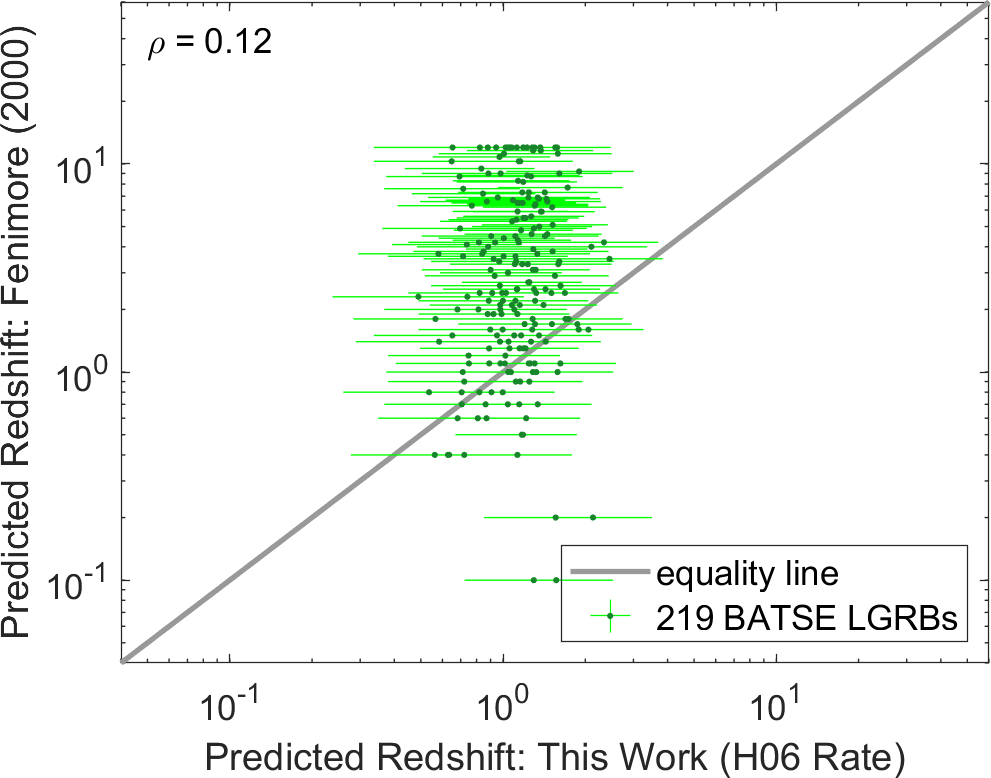} & \includegraphics[width=0.235\textwidth]{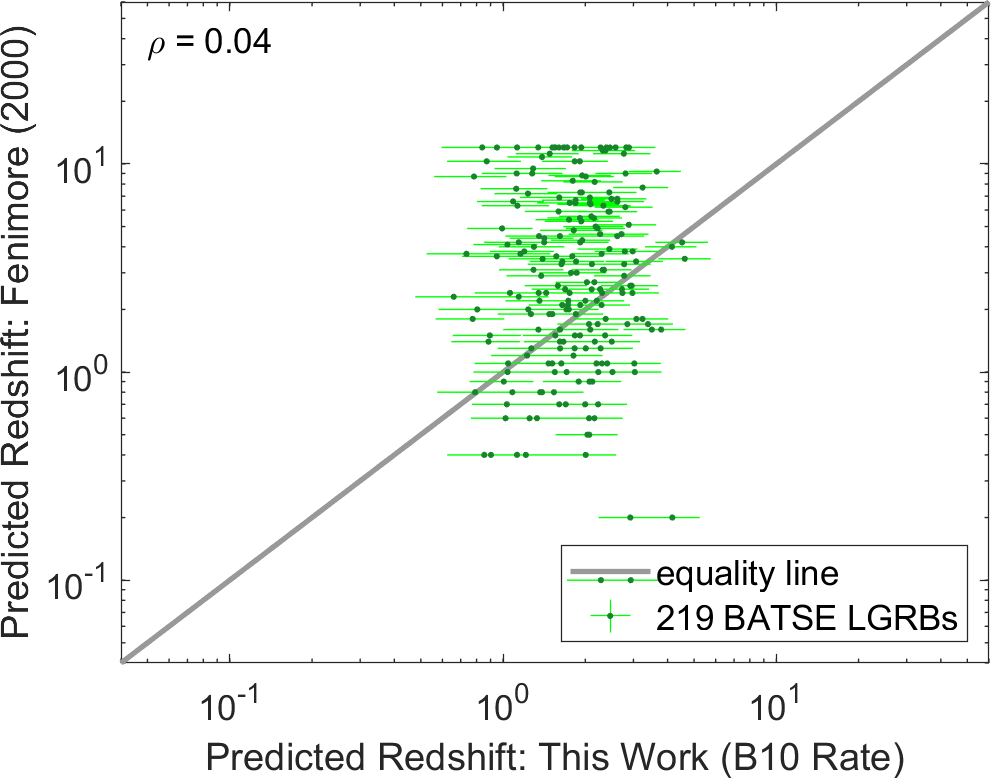} & \includegraphics[width=0.235\textwidth]{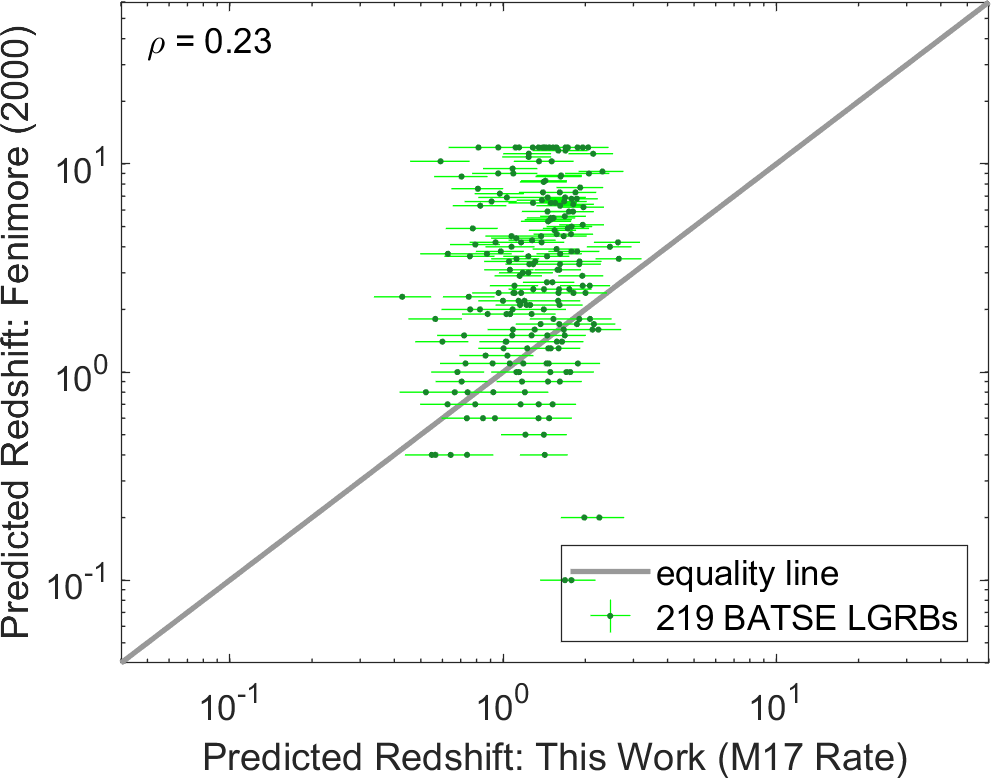} & \includegraphics[width=0.235\textwidth]{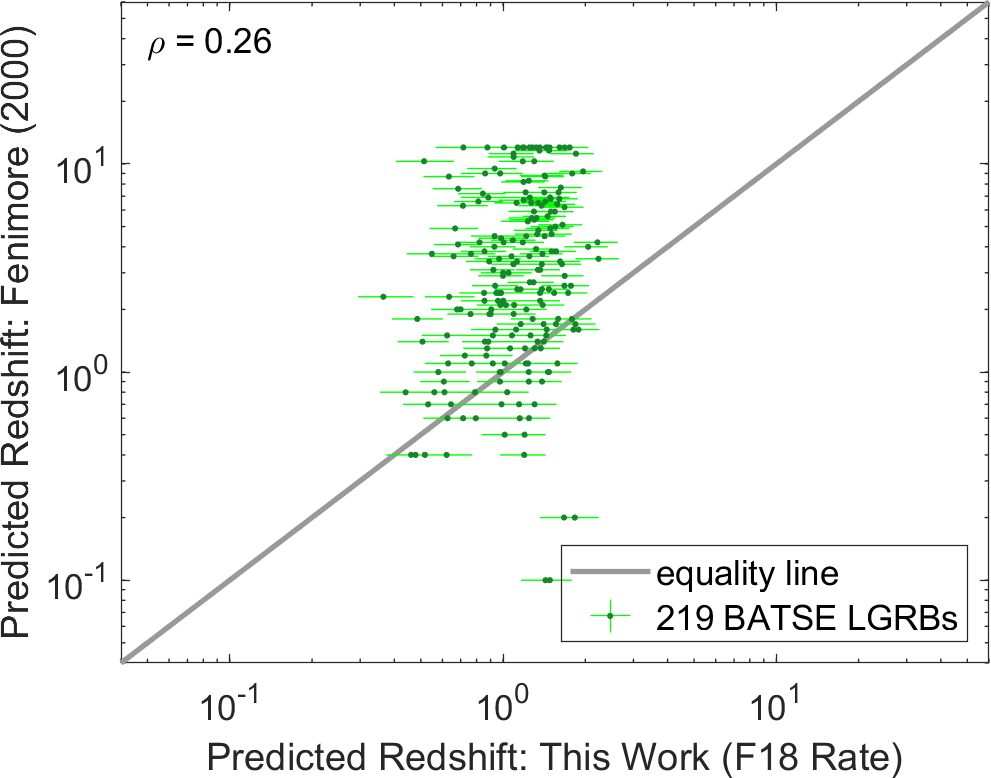} \\
            \includegraphics[width=0.235\textwidth]{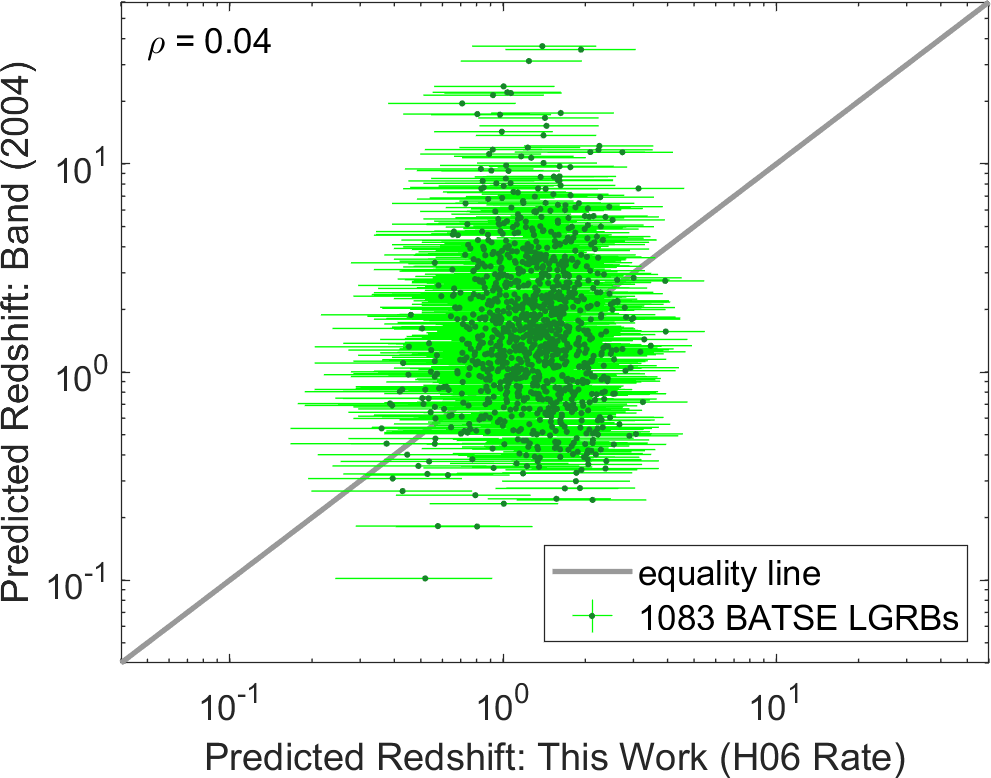} & \includegraphics[width=0.235\textwidth]{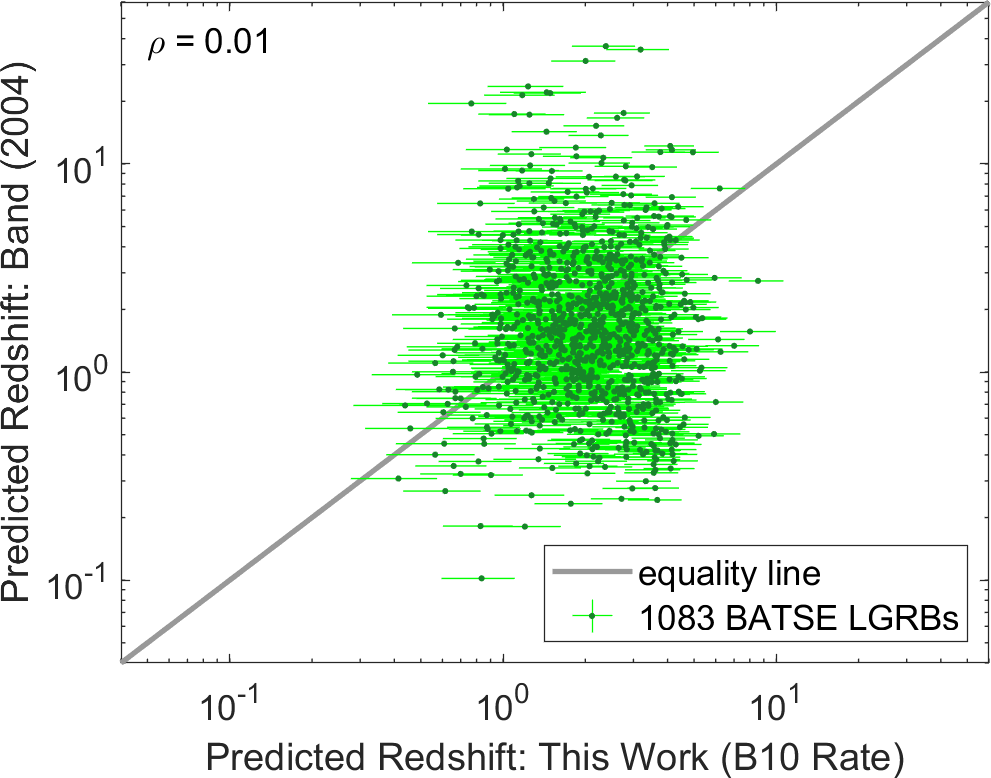} & \includegraphics[width=0.235\textwidth]{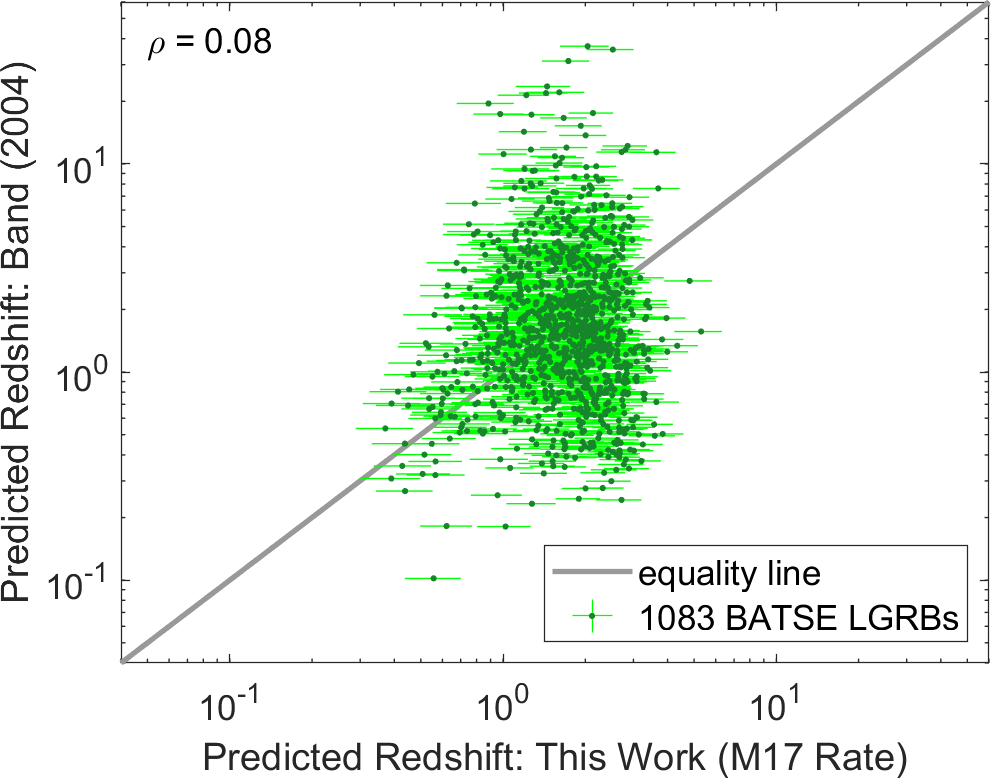} & \includegraphics[width=0.235\textwidth]{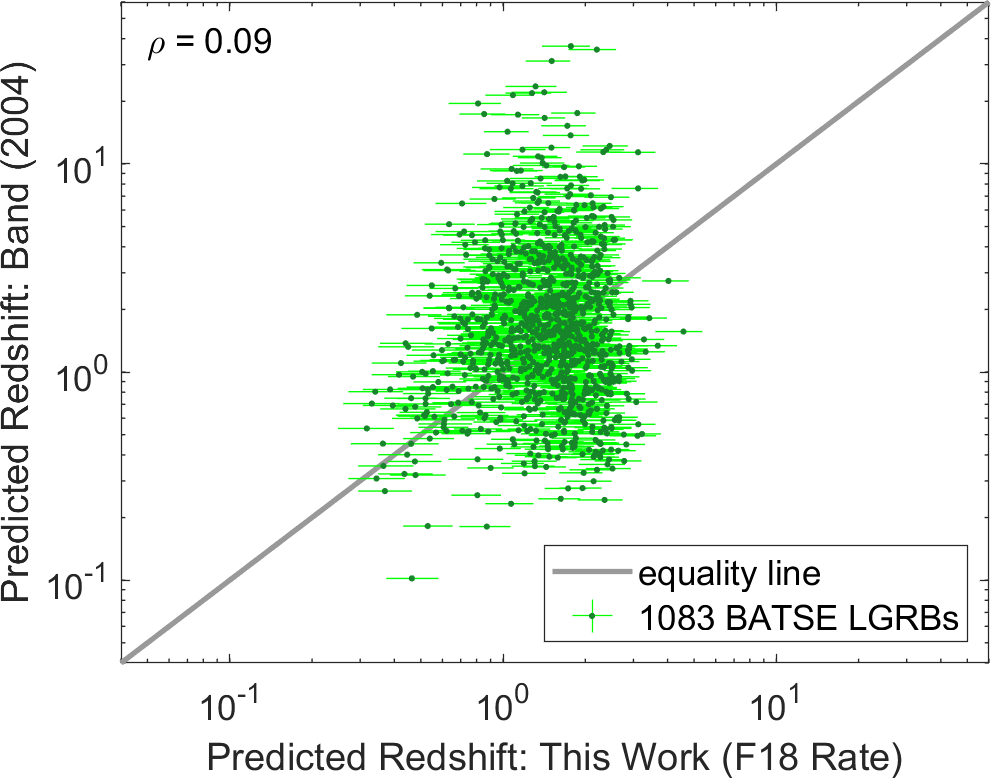} \\
            \includegraphics[width=0.235\textwidth]{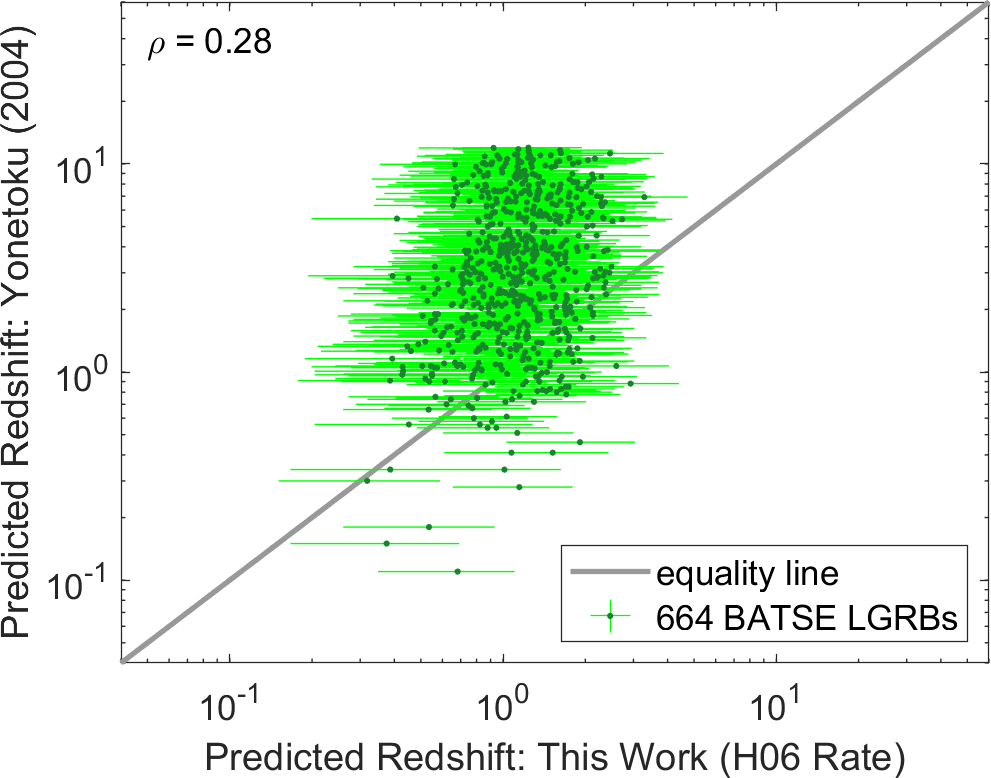} & \includegraphics[width=0.235\textwidth]{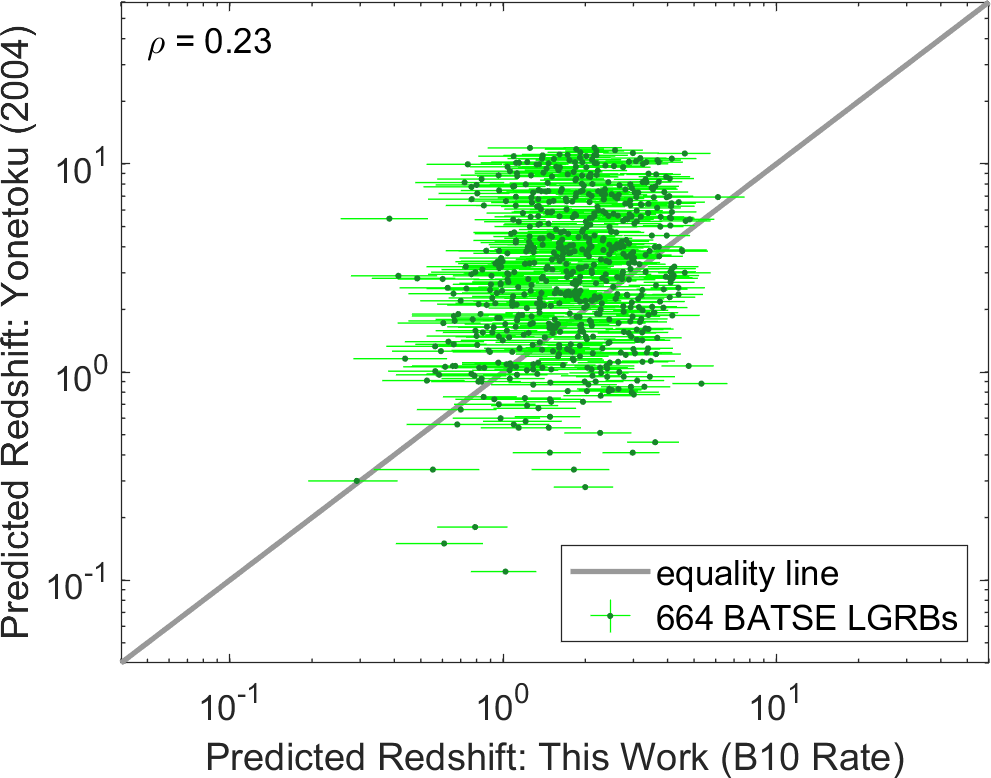} & \includegraphics[width=0.235\textwidth]{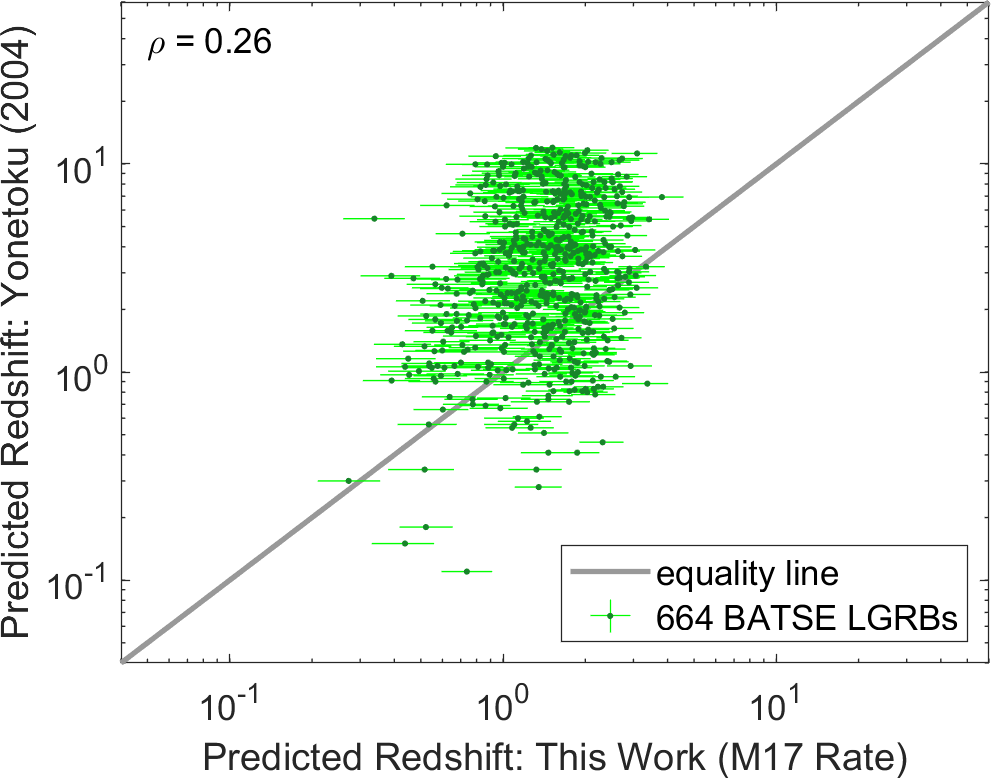} & \includegraphics[width=0.235\textwidth]{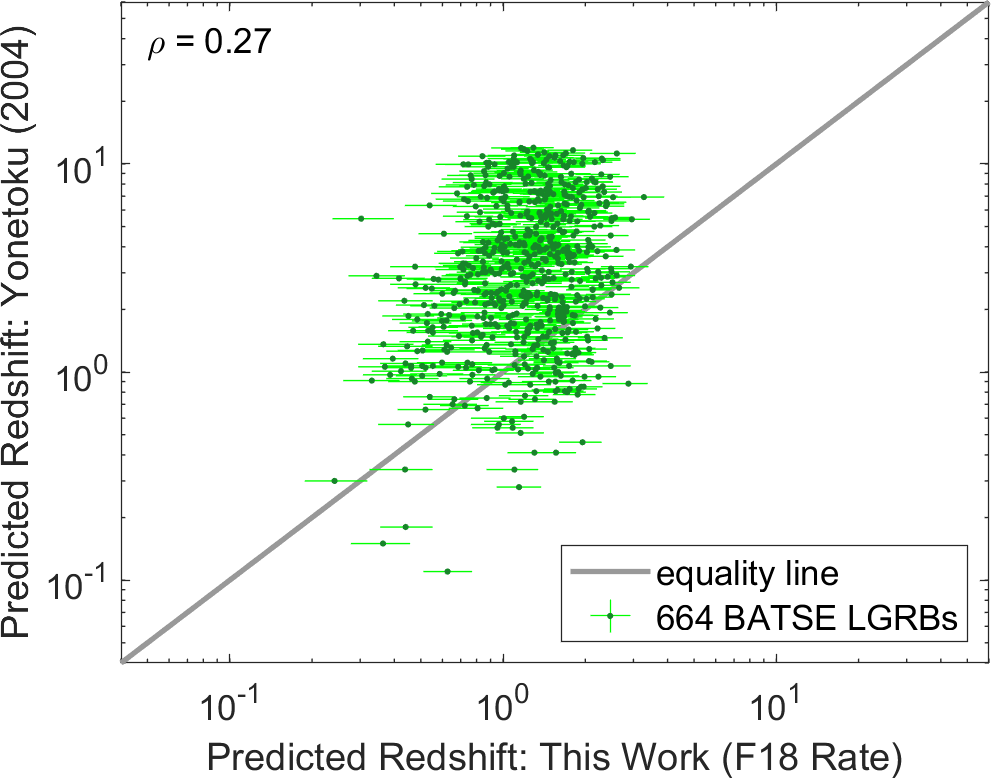} \\
        \end{tabular}
        \caption{
        A comparison of the expected redshifts of BATSE LGRBs under the four different cosmic LGRB rate densities assumptions \eqref{eq:mz} with the BATSE redshift estimates from previous works, based on the correlation between lightcurve-variability and the peak luminosity by \citet{fenimore2000redshifts}, the lag-luminosity relation by \citet{band2004gamma}, and the correlation between the peak luminosity and the spectral peak energy by \citet{yonetoku2004gamma}. On each plot, the Spearman's correlation coefficient of the two sets of expected redshifts is also reported. The error bars represent the $90\%$ predictions intervals for each predicted redshift in this work.\label{fig:redshiftComparisonWithOldWorks}}
    \end{figure*}

    The larger the variances of redshift-related terms in these equations are (compared to the variances of rest-frame LGRB attributes), the more the observer-frame LGRB properties will be indicative of the redshifts of individual events. For example, the LGRB rate density of \citetalias{fermi2018gamma} results in the largest ratios of the variances of redshift-related terms \eqref{eq:obsIntMap} to the variances of the intrinsic BATSE LGRB attributes. This, in turn, leads to the least uncertain (but not necessarily the most accurate) individual redshift predictions under the rate density assumptions of \citetalias{fermi2018gamma} among all rate densities considered in this study. This is also evidenced in the plots of Figure \ref{fig:expectedRedshift} and \ref{fig:redshiftComparisonWithOldWorks} by the relatively tighter prediction intervals (i.e., error bars) for the redshift estimates based on the LGRB rate density of \citetalias{fermi2018gamma}.\newpar

    \begin{figure*}[tphb]
        \centering
        \begin{tabular}{ccc}
            \includegraphics[width=0.316\textwidth]{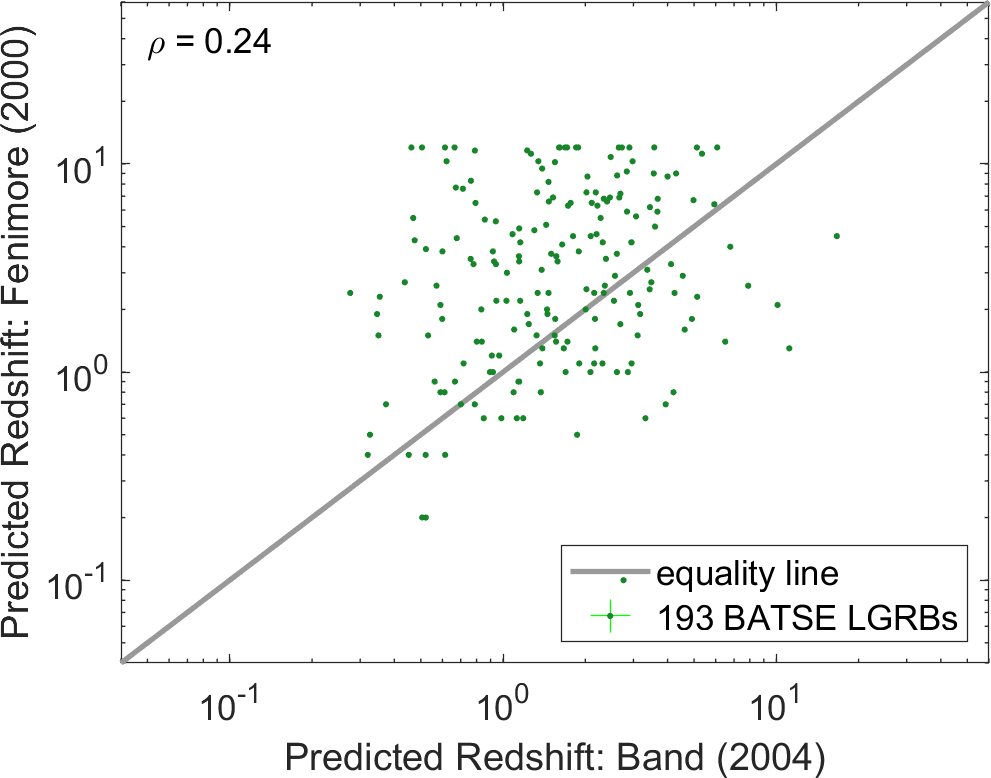} &
            \includegraphics[width=0.316\textwidth]{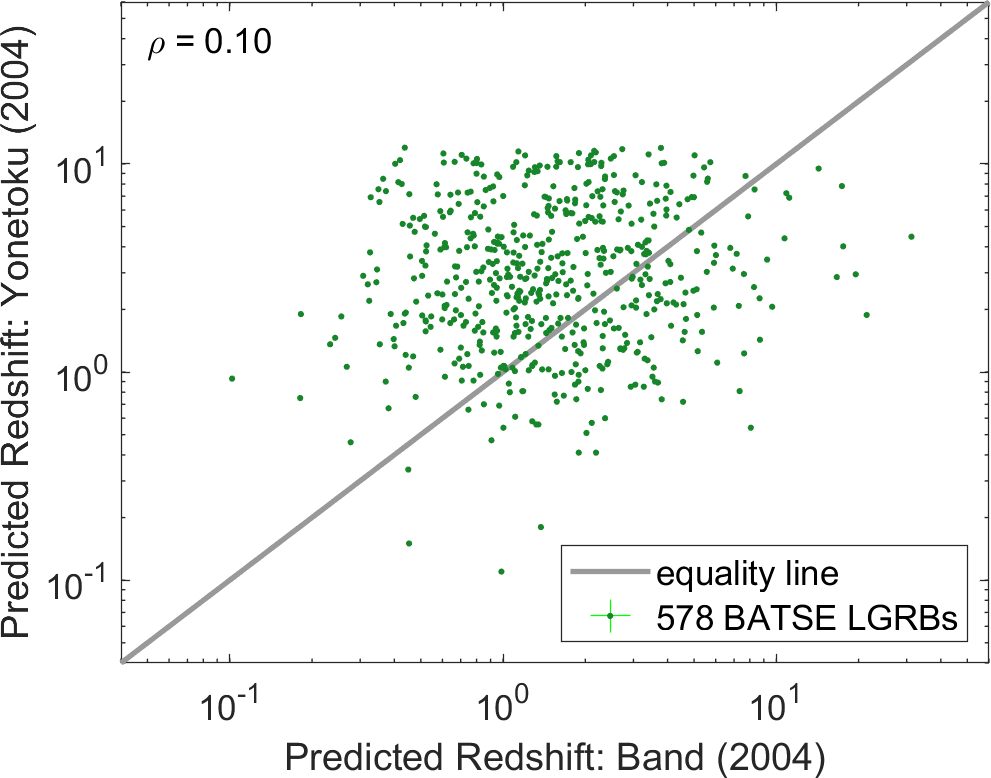} &
            \includegraphics[width=0.316\textwidth]{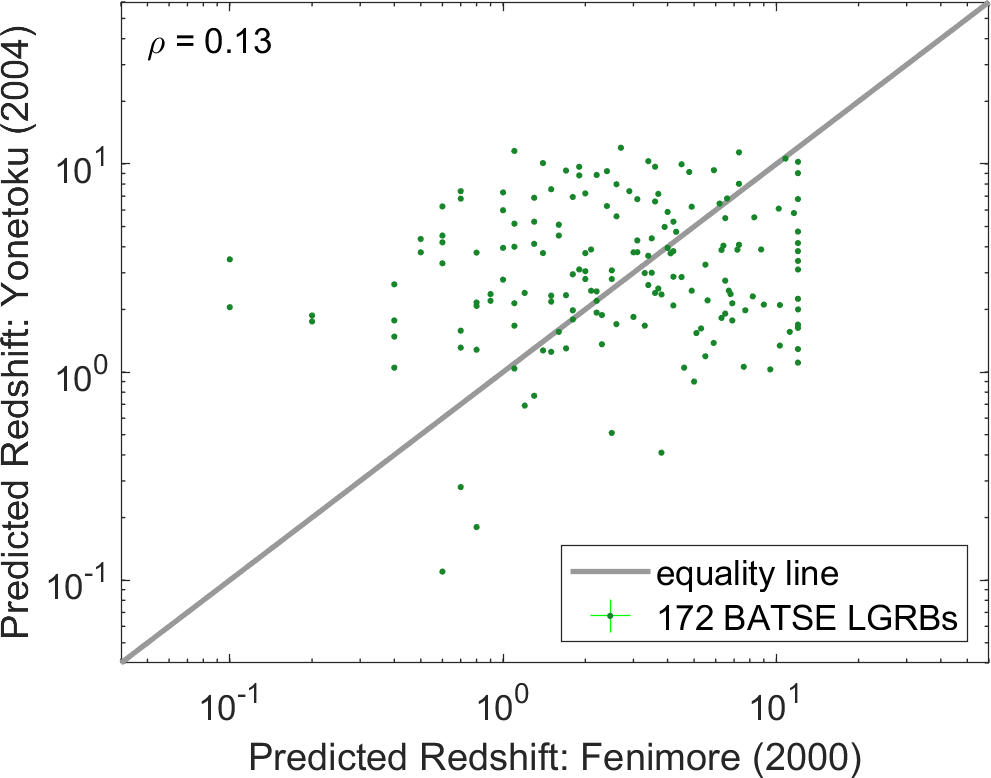} \\
        \end{tabular}
        \caption{A comparison of the predicted redshifts of BATSE GRBs based on the correlation between lightcurve-variability and the peak luminosity by \citet{fenimore2000redshifts}, the lag-luminosity relation by \citet{band2004gamma}, and the correlation between the peak luminosity and the spectral peak energy by \citet{yonetoku2004gamma} with each other. On each plot, the Pearson's correlation coefficient of the two sets of predicted redshifts is also reported.\label{fig:oldWorkComparison}}
    \end{figure*}

    Several previous studies have attempted to estimate the unknown redshifts of BATSE LGRBs via some of the phenomenological gamma-ray correlations. For example, \citet{fenimore2000redshifts} used the apparent correlation between the isotropic peak luminosity of GRBs and the temporal variability of their lightcurves to estimate redshifts of 220 BATSE GRBs. \citet{band2004gamma} used the apparent correlation between the spectral lag and the peak luminosity of GRBs to estimate the unknown redshifts of 1194 BATSE events. Similarly, \citet{yonetoku2004gamma} used the apparent observed correlation between the isotropic peak luminosity and the intrinsic spectral peak energies of GRBs to estimate redshifts of 689 BATSE GRBs.\newpar

    We compare our redshift estimates under different LGRB rate density assumptions to the predictions of each of the aforementioned works in Figure \ref{fig:redshiftComparisonWithOldWorks}. None of these three previous independent redshift estimates based on the high-energy correlations appear to agree with our predictions. These three independent estimates are also highly inconsistent with each other as shown in Figure \ref{fig:oldWorkComparison}.\newpar

    The observed inconsistencies of the previous independent redshift estimates of BATSE GRBs with each other, as well as with the predictions of this work may be explained by the fact that the prompt gamma-ray correlations used in the works of \citet{fenimore2000redshifts}, \citet{band2004gamma}, and \citet{yonetoku2004gamma} were constructed from a small sample of heterogeneously collected GRB events, and that the observed phenomenological relations are likely severely affected by sample incompleteness. Despite the discrepancies in the redshift estimates of individual BATSE LGRBs, our predictions corroborate the findings of some previous studies \citep[e.g.,][]{ashcraft2007there, hakkila2009gamma} in that the probability of the existence of very high-redshift LGRBs in BATSE catalog is negligible.
    \newpar

     As illustrated in Figure \ref{fig:redshiftComparisonWithOldWorks}, the fact that the redshift estimates of \citet{yonetoku2004gamma} show the least disagreement with our predictions among all previous attempts, can be explained by noting the relative similarity of the assumptions in \citet{yonetoku2004gamma} to infer the redshifts with the findings of this work: \citet{yonetoku2004gamma} infer the redshifts based on the assumption of the existence of a tight positive correlation between the intrinsic spectral peak energy and the peak luminosity of LGRBs. Our modeling approach confirms the existence of such a positive correlation (see Table \ref{tab:paraPostStat}), albeit with much higher dispersion and less strengths. The disparity in the predicted strength and regression-slope of this correlation can reasonably explain the non-negligible disagreement between our predictions and those of \citet{yonetoku2004gamma}.

    \begin{figure*}[tphb]
        \centering
        \begin{tabular}{cccc}
            \includegraphics[width=0.235\textwidth]{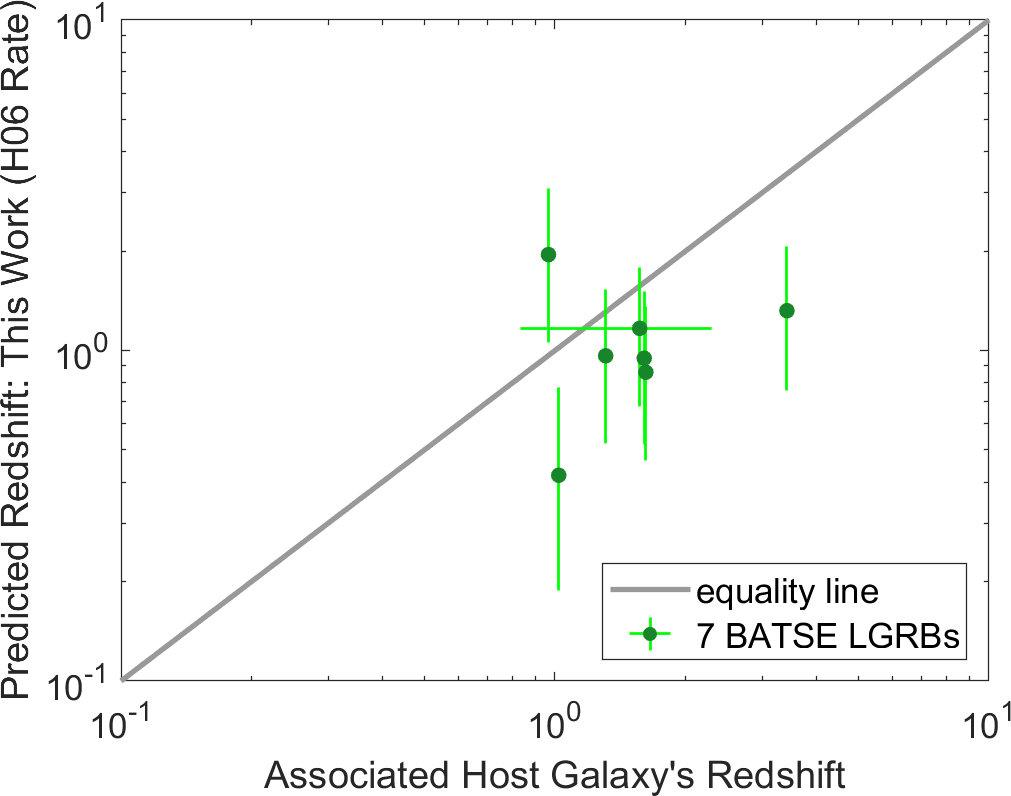} &
            \includegraphics[width=0.235\textwidth]{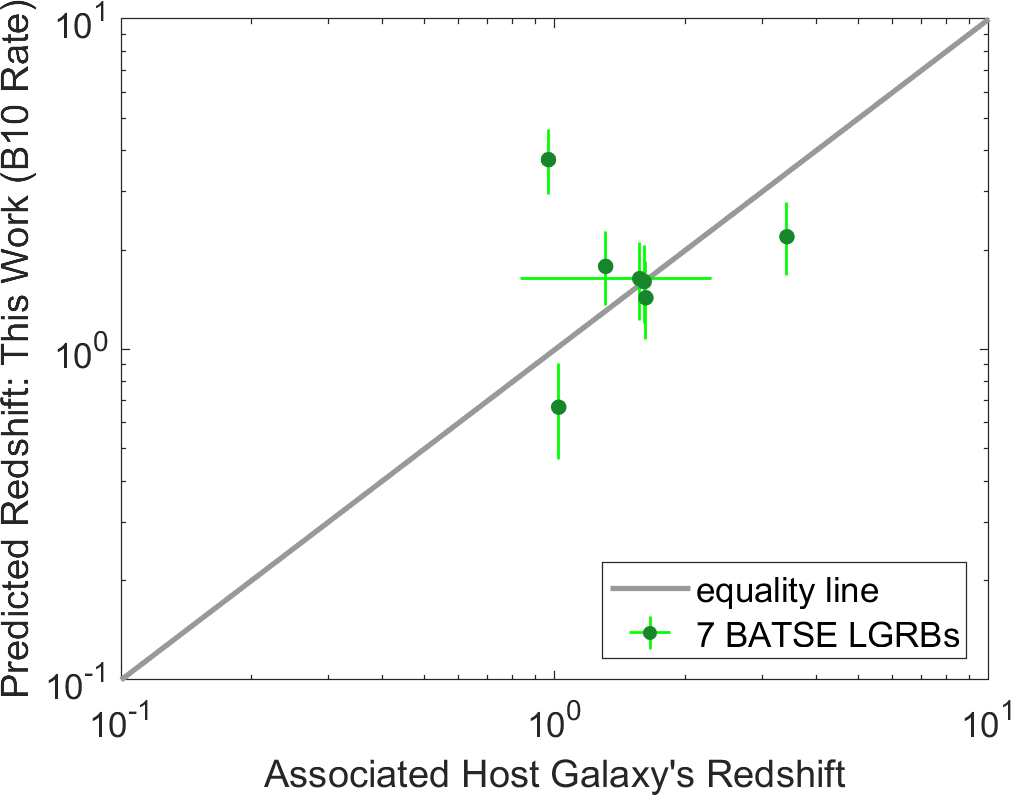} &
            \includegraphics[width=0.235\textwidth]{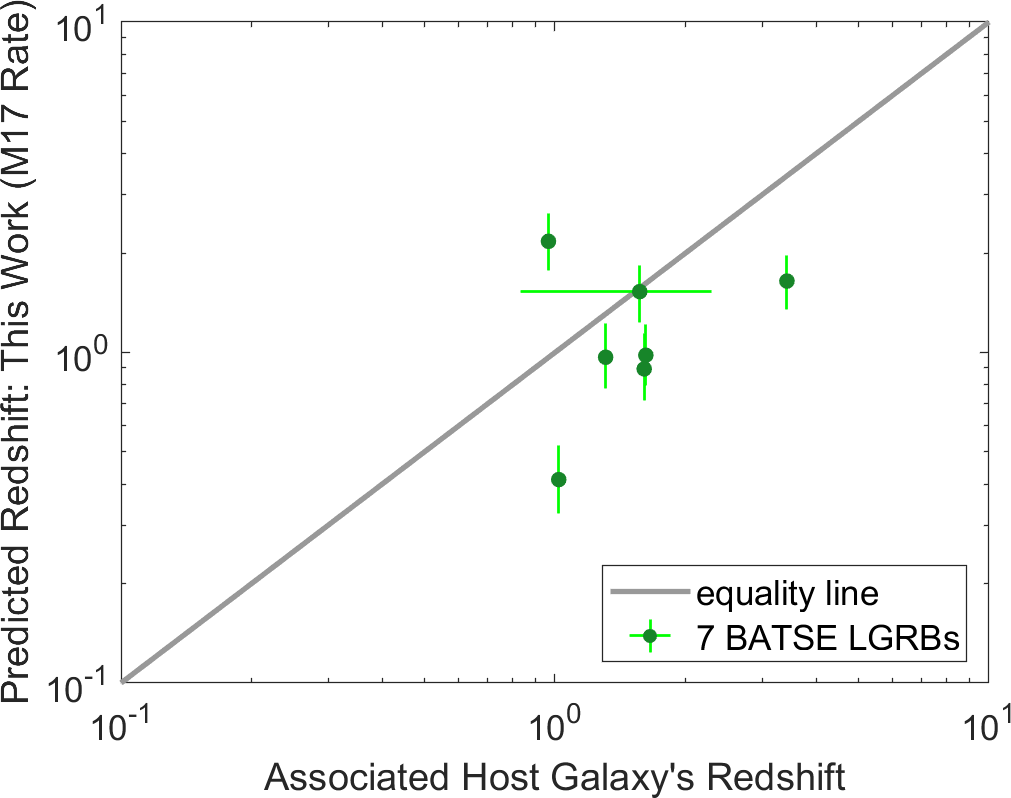} &
            \includegraphics[width=0.235\textwidth]{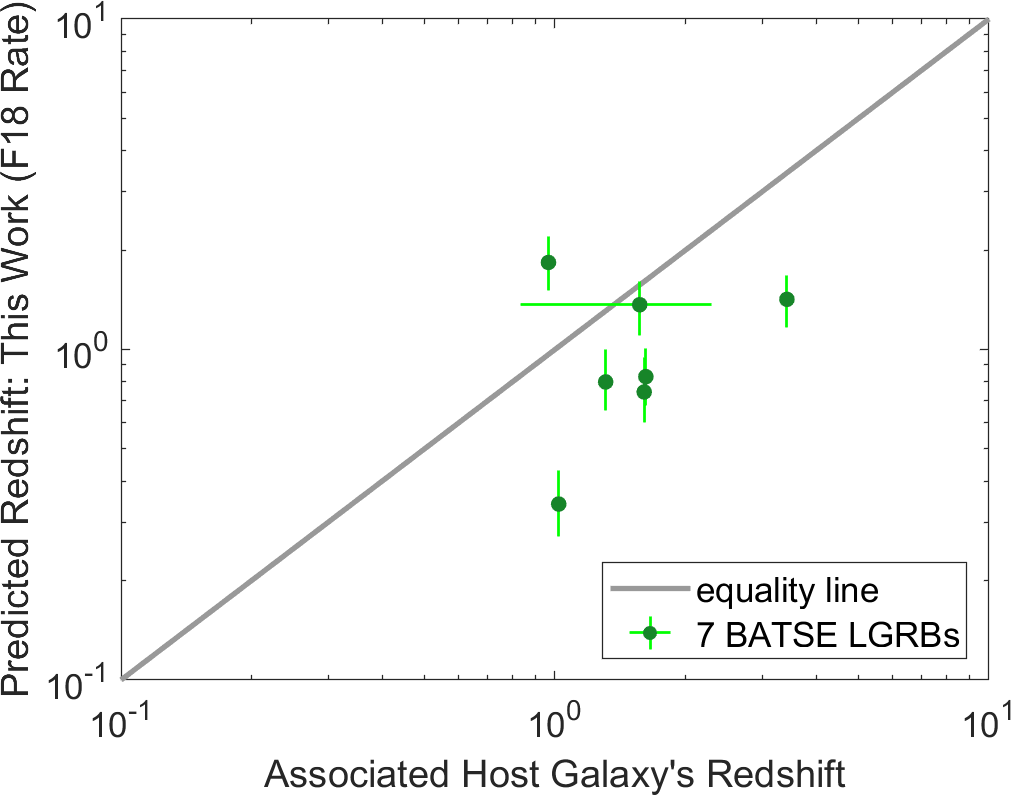} \\
        \end{tabular}
        \caption{A comparison of the predicted redshifts of a set of BATSE LGRBs with their reported measured redshifts in the literature, for the four LGRB rate density assumptions of \citetalias{hopkins2006normalization}, and \citetalias{butler2010cosmic}, \citetalias{madau2017radiation}, \citetalias{fermi2018gamma}. The error bars represent the 90\% uncertainty intervals for the predicted redshifts and the 100\% uncertainty intervals for the measured redshifts (where available).\label{fig:knownRedshifts}}
    \end{figure*}

    Finally, we compare our predictions with the measured redshifts of BATSE events, where available. Out of 2702 BATSE LGRBs, only less than a dozen have measured redshifts through association with their potential host galaxies. In Figure \ref{fig:knownRedshifts}, we compare the reported redshifts of 7 such BATSE events (that also exist in our catalog) with their corresponding predicted redshifts in this work. Among all LGRB rate densities models considered, the predicted redshifts based on the LGRB rate of \citetalias{butler2010cosmic} appear to show the highest level of consistency with the measured redshifts. This leads us to cautiously conclude that {\bf the LGRB formation rate may not exactly trace the star formation rate in the distant universe, corroborating the previous finding of \citetalias{butler2010cosmic}, \citet{shahmoradi2013multivariate}, and \citet{shahmoradi2015short}}.\newpar

    A remarkable modeling assumption in our work is the redshift-independence of the proposed model for the four main gamma-ray properties of cosmic LGRBs (the term $\mintlgrb$ in \eqref{eq:modelint}). This modeling assumption is similar to the assumptions of \citet{shahmoradi2013multivariate, shahmoradi2015short} and follows the findings of \citetalias{butler2010cosmic} who, based on a multivariate analysis of a large sample of Swift LGRBs with and without redshifts, reject the hypothesis of an evolving redshift-dependent luminosity function for LGRBs. This is in contrast to the findings of \citet{petrosian2015cosmological, pescalli2016rate, yu2015unexpectedly, tsvetkova2017konus} who model the cosmic rates of LGRBs via a redshift-dependent luminosity function.\newpar

    Regardless of the validity of the aforementioned assumption in our modeling, we note that it is practically impossible to infer any redshift-dependence of the energetics and the luminosity function of LGRBs solely via BATSE LGRBs data. This limitation primarily results from the complete lack of knowledge of the redshifts of BATSE LGRBs. This missing knowledge leads to highly degenerate parameter space for models that consider the possibility of a redshift evolution of the prompt emission properties of LGRBs.\newpar

    While being a remote possibility, one of the potential caveats of our presented redshift estimates is that, if an SGRB has been mistakenly classified as an LGRB in our catalog of 1366 by our classification method described in \ref{sec:methods:data}, then its estimated redshift may not be accurate. Also, this work did not take into account the potential effects of the GRBs' jet beaming angle. A recent study by \citet{lazzati2013photospheric} finds that the different orientations of the GRB jet axis with respect to the observer could partially explain the observed LGRB brightness-hardness type relations. Such an effect could potentially lead to more uncertainty in the predicted redshifts and perhaps could explain the lack of a complete perfect agreement between the known redshifts of a handful of BATSE LGRBs and their corresponding predicted redshifts in this work, as illustrated by the plots of Figure \ref{fig:knownRedshifts}. These are among improvements that could be made in the future to our mathematically-rigorous, purely-probabilistic, bias-aware approach to estimating or further constraining the unknown redshifts of GRBs in the currently-available and future GRB catalogs.

\acknowledgments

This work would have not been accomplished without the vast time and effort spent by many scientists and engineers who designed, built and launched the Compton Gamma-Ray Observatory and were involved in the analysis of GRB data from BATSE Large Area Detectors.\newpar

\bibliographystyle{aasjournal}
\bibliography{../../../../libtex/all}

\appendix

\begin{center}
\begin{longtable*}{ccccccccc}
        \caption[]{BATSE 1366 LGRB Redshift Estimates \label{tab:redshiftCatalog}} \\
        \hline \hline \\[0.5ex]
        Trigger &
        $\mu_{\mathrm{Z,H06}}$ & PI$_{\mathrm{90\%}}$ &
        $\mu_{\mathrm{Z,B10}}$ & PI$_{\mathrm{90\%}}$ &
        $\mu_{\mathrm{Z,M17}}$ & PI$_{\mathrm{90\%}}$ &
        $\mu_{\mathrm{Z,F18}}$ & PI$_{\mathrm{90\%}}$
        \\[0.5ex] \hline \\[0.5ex]
        \endfirsthead
        %
        \multicolumn{9}{c}{{\tablename} \thetable{} -- Continued} \\[0.5ex]
        \hline \hline \\[0.5ex]
        Trigger &
        $\mu_{\mathrm{Z,H06}}$ & PI$_{\mathrm{90\%}}$ &
        $\mu_{\mathrm{Z,B10}}$ & PI$_{\mathrm{90\%}}$ &
        $\mu_{\mathrm{Z,M17}}$ & PI$_{\mathrm{90\%}}$ &
        $\mu_{\mathrm{Z,F18}}$ & PI$_{\mathrm{90\%}}$
        \\[0.5ex] \hline \\[0.5ex]
        \endhead
        %
        \\[0.5ex] \hline \\[0.5ex]
        \multicolumn{9}{r}{{Continued on Next Page \ldots}} \\[0.5ex]
        \endfoot
        \\[0.5ex] \hline \hline
        \endlastfoot
        105 & 0.46 & $[0.22 , 0.81]$ & 0.59 & $[0.39 , 0.83]$ & 0.56 & $[0.43 , 0.71]$ & 0.49 & $[0.37 , 0.60]$ \\ [0.5ex]
107 & 1.61 & $[0.88 , 2.56]$ & 2.70 & $[1.98 , 3.55]$ & 2.55 & $[1.92 , 3.15]$ & 2.30 & $[1.68 , 2.81]$ \\ [0.5ex]
109 & 1.69 & $[0.93 , 2.71]$ & 3.24 & $[2.53 , 4.02]$ & 1.90 & $[1.57 , 2.31]$ & 1.59 & $[1.32 , 1.90]$ \\ [0.5ex]
110 & 2.48 & $[1.31 , 3.87]$ & 4.47 & $[3.53 , 5.57]$ & 3.33 & $[2.61 , 3.98]$ & 2.93 & $[2.24 , 3.47]$ \\ [0.5ex]
111 & 2.46 & $[1.29 , 3.87]$ & 4.62 & $[3.67 , 5.75]$ & 3.09 & $[2.50 , 3.67]$ & 2.61 & $[2.08 , 3.06]$ \\ [0.5ex]
114 & 2.04 & $[1.07 , 3.27]$ & 3.98 & $[3.08 , 5.00]$ & 2.75 & $[2.16 , 3.31]$ & 2.34 & $[1.82 , 2.79]$ \\ [0.5ex]
121 & 2.29 & $[1.22 , 3.61]$ & 4.18 & $[3.35 , 5.13]$ & 2.72 & $[2.22 , 3.23]$ & 2.30 & $[1.88 , 2.71]$ \\ [0.5ex]
130 & 1.62 & $[0.90 , 2.59]$ & 3.02 & $[2.34 , 3.78]$ & 1.88 & $[1.55 , 2.27]$ & 1.58 & $[1.31 , 1.88]$ \\
\end{longtable*}
\end{center}
{Notes: The column denoted by `Trigger' contains the trigger IDs of BATSE LGRBs. The columns denoted by $\mu_{H06}$, $\mu_{B10}$, $\mu_{M17}$, $\mu_{F18}$ contain the predicted mean redshifts of individual BATSE LGRBs, based on the four LGRB rate model assumptions (\citetalias{hopkins2006normalization}, \citetalias{butler2010cosmic}, \citetalias{madau2017radiation}, \citetalias{fermi2018gamma}) considered in this work as given by \eqref{eq:mz}, \eqref{eq:pz}, \eqref{eq:mzm}, \eqref{eq:pzm}. The rest of the columns denoted by `PI$_{90\%}$' contain the lower and upper bounds of the 90\% Prediction Intervals (i.e., the most probable ranges) for the unknown redshifts of individual BATSE LGRBs. This table as well as the full probability density functions of the redshifts of individual BATSE LGRBs are available for download at the permanent GitHub repository of this work.}

\newpage

\end{document}